\documentclass[aps,prb,reprint,superscriptaddress,longbibliography]{revtex4-1}
\usepackage{amsmath}
\usepackage{graphicx}
\usepackage{amsfonts}
\usepackage{color}
\usepackage{upgreek}
\usepackage{bm}
\usepackage{lipsum}
\usepackage{comment}
\usepackage{enumerate}
\usepackage{dsfont}

\usepackage[table]{xcolor}
\usepackage{textcomp}
\usepackage{amsmath}
\usepackage{amssymb}
\usepackage{graphicx} 
\usepackage{pgf}
\usepackage{epstopdf}
\usepackage{braket}
\usepackage{mathtools}
\usepackage{times}

\usepackage[retainorgcmds]{IEEEtrantools}
\newcommand{\bea}{\begin{eqnarray}}
\newcommand{\eea}{\end{eqnarray}}

\newcommand{\Tr}{\mathrm{Tr}}
\renewcommand{\Im}{\,\mathrm{Im}}

\usepackage[colorlinks,linkcolor=blue,urlcolor=blue,citecolor=blue]{hyperref}

\begin{document}

\title{Role of Bath-Induced Many-Body Interactions in the Dissipative Phases of the Su-Schrieffer-Heeger Model} 


\author{Brett Min}
\email{brett.min@mail.utoronto.ca}
\affiliation{Department of Physics and Centre for Quantum Information and Quantum Control, University of Toronto, 60 Saint George St., Toronto, Ontario, M5S 1A7, Canada}

\author{Kartiek Agarwal}
\affiliation{Material Science Division, Argonne National Laboratory, Argonne, IL 60439, USA}
\affiliation{Department of Physics, McGill University, Montréal, Québec, Canada H3A 2T8}

\author{Dvira Segal}
\email{dvira.segal@utoronto.ca}
\affiliation{Department of Chemistry
University of Toronto, 80 Saint George St., Toronto, Ontario, M5S 3H6, Canada}

\affiliation{Department of Physics and Centre for Quantum Information and Quantum Control, University of Toronto, 60 Saint George St., Toronto, Ontario, M5S 1A7, Canada}

\begin{abstract}
The Su-Schrieffer-Heeger chain is a prototype example of a symmetry-protected topological insulator. Coupling it non-perturbatively to local thermal environments, either through the intercell or the intracell fermion tunneling elements, modifies  the topological window. 
To understand this effect, we employ the recently developed reaction-coordinate polaron transform (RCPT) method,
which allows treating system-bath interactions at arbitrary strengths. The effective system Hamiltonian, which is obtained via the RCPT, exposes the impact of the baths on the SSH chain through renormalization of tunneling elements and the generation of many-body interaction terms. By performing exact diagonalization and computing the ensemble geometric phase, a topological invariant applicable even to systems at finite temperature, we  distinguish the trivial band insulator (BI) from the topological insulator (TI) phases. Furthermore, through the RCPT mapping, we are able to pinpoint the main mechanism behind the extension of the parameter space for the TI or the BI phases (depending on the coupling scheme, intracell or intercell), which is the bath-induced, dimerized, many-body interaction. We also study the effect of on-site staggered potentials on the SSH phase diagram, and discuss extensions of our method to higher dimensions.
\end{abstract}
\maketitle

\date{\today}

\section{Introduction}
Controlling and engineering Hamiltonians with exotic properties is of prime interest in modern condensed matter physics and materials science. In particular, Hamiltonians exhibiting topological phases of matter have attracted significant attention. This interest is largely focused on 
topological superconductors, which support quasi-particle excitations with non-Abelian exchange statistics. 
By enabling non-local storage of quantum information, these materials may see applications in fault-tolerant quantum computation  \cite{Kitaev_2001,Alicea_2012}.
Although theoretical proposals for devising such systems are well-established \cite{Lutchyn_2010,Oreg_2010,Sau_2010}, the experimental realization and manipulation of these systems \cite{Prada_2012,Mourik_2012,Churchill_2013,Finck_2013,Yazdani_2014,Albrecht_2016,Deng_2016,Suominen_2017,Nichele_2017,Marcuss_2018,Deng_2018,Vaitiekėnas_2020,Pan_2020,Sarma_2021} remain a challenge. Developing new theoretical approaches to make these systems more robust, easier to create and manipulate, or capable of undergoing a topological phase transition from a trivial phase through an external intervention is a continuous pursuit.  

One popular method of engineering Hamiltonians with a desired structure is known as \textit{Floquet} engineering, where periodic driving of the material enhances certain symmetries and/or suppresses undesired terms. Although persistent driving leads to eventual heating of the system destroying any useful or interesting order, driving at high frequencies~\cite{Mori_2016,Abanin_2017} can slow down heating to times exponentially long in the frequency of the drive. 
Indeed, there have been a number of theoretical studies in employing Floquet engineering \cite{Martin_2020,Agarwal_2020,Min_2022,Ling_2024} to enhance symmetries and/or to make topological modes in these systems more robust, and understand the effect of noise on such techniques~\cite{Martin_2022}. 

A potentially greater challenge than heating is that the engineered Floquet eigenstates are inherently non-equilibrium states, and it is usually not easy to engineer appropriate baths that will eventually allow the system to relax into these Floquet states~\cite{Refaelandco}. In other words, in this approach, one does not have  access to a stable ground state or an ensemble of ``low energy'' states. A more elegant approach, 
developed in the field of cavity quantum materials \cite{Trif_2019,Nie_2020,Huang_2020},
allows circumventing both these issues:
recent experimental developments in this field have realized setups where a many-body system and the vacuum field inside a cavity interact strongly \cite{Scalari_2012,Chikkaraddy_2016,Lamata_2019}. This strong light-matter interaction can potentially \emph{both} coherently modify phase boundaries and lead to realization of novel phases \emph{and} serve as a thermal bath and drive the system to thermal equilibrium described by a Gibbs' ensemble with a modified system Hamiltonian. Coupling to external degrees of freedom thus provides an alternative route to engineer and control many-body systems, an approach that does not require an external time-dependent driving \cite{Ashida_2020,Ashida_2021,Ashida_2023,Masuki_2023,Masuki_2024}.

On a separate note, there has been a recent surge of interest in extending the concept of the usual topological invariant, which traditionally applies to noninteracting systems at zero temperature, to interacting, open, and nonzero temperature systems \cite{Budich_2015,Grusdt_2017,Zheng_2018,Wawer_2021,He_2022,Gao_2023,Mera_2017,Unanyan_2020,Xing_2021,Kuno_2019,Zhou_2023,Melo_2023,Yahyavi_2018,Zhang_2018,Gneiting_2022,Rivas_2013,Lieu_2020,Wang_2013}. One of the major reasons for this interest is the potential to control and engineer topological phases of matter by enabling systems exhibiting these phases to interact with the external environment. It is generally believed that when such systems interact with an external thermal 
environment, their topological phases are lost. However, by carefully engineering the nature of the interaction between the system and its surroundings, it is possible to induce a topological phase transition, from a trivial phase to a non-trivial one \cite{Zhang_2022,Linzner_2016,Beck_2021,Nair_2023,Nie_2021,McDonnell_2022,Bardyn_2013,Takata_2018}. These systems are usually studied in the context of open topological systems using non-Hermitian or Lindbladian approaches \cite{Gong_2018,Dangel_2018,Nava_2023,ZhangJH_2023,Pereira_2024}.

With new experimental setups achieving strong system-environment couplings, and development of working  definitions for topological invariants, applicable beyond zero temperature, it is timely to 
study 
systems that are non-perturbatively coupled to a thermal environment and exhibit topological phases. 
In particular, it is of a great interest to show that the interaction of a an electronic system with the environment could lead to a topological phase transition, or the enhancement of the original topological phase. 

\begin{figure}[htbp]
\fontsize{6}{10}\selectfont 
\centering
\includegraphics[width=0.9\columnwidth]{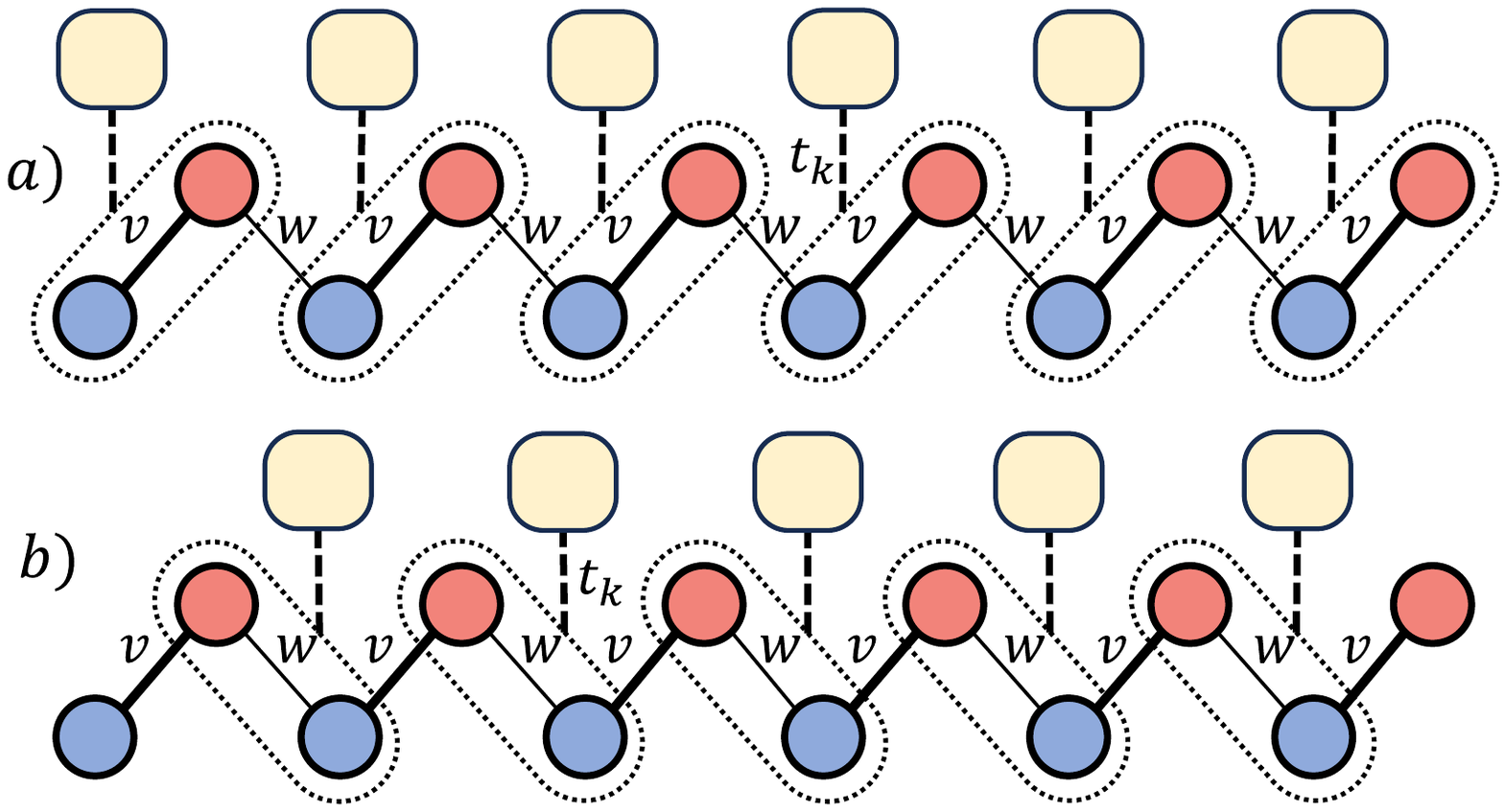}
\caption{Two dissipative SSH models. (a) Intracell coupling to local baths. Here,  fermion hopping matrix elements, with amplitude $v$, are coupled to independent bosonic baths. 
(b) Intercell coupling to local baths. In this scheme, the fermionic hopping matrix elements of amplitude $w$ are coupled to independent bosonic baths. 
The blue (red) sites in the scheme correspond to the A (B) sites.
The yellow squares depict local heat baths, assumed to be identical and maintained at the same temperature.
}
\label{fig:figure 1}
\end{figure}

In this paper, we study the phases of the dissipative
Su-Schrieffer-Heeger (SSH) model at nonzero temperatures. The SSH model describes spinless fermions on a one-dimensional chain with a two-site unit cell, at half-filling \cite{Su_1979}. 
Electrons hopping along the chain are coupled to local heat baths in two different ways, depicted in Fig. \ref{fig:figure 1}.
The first is the \textit{intracell} 
coupling scheme, where the intracell hopping Hamiltonian couples to local bosonic baths. The second is the \textit{intercell} coupling scheme, where the unit-cell boundary hopping Hamiltonian couples to local bosonic baths.
These dissipative scenarios were recently studied using the quantum Monte Carlo (QMC) and cluster perturbation theory (CPT) approaches \cite{Pavan_2024}, where based on simulations, the intracell coupling scheme was found to be detrimental to the topological phase, while the intercell coupling scheme was found to be advantageous. 

With the objective to bring a 
deep understanding to how intracell and intercell couplings to thermal environments allow engineering of topological phases, in this work we derive the phase diagram of the dissipative SSH model using a powerful {\it analytical tool}. This technique holds even at strong electron-phonon coupling. It allows the construction of an effective system Hamiltonian that is \textit{Hermitian}. This effective Hamiltoian exposes directly the impact of the baths on the SSH chain through renormalization of hopping terms, suppression of site energies (in the staggered case), and most importantly, the generation of a many-body bath-mediated fermion-fermion term. We then apply a numerically exact diagonalization technique to the effective (dissipation-dressed) Hamiltonian, and study its phase with either intracell or intercell hopping terms, coupled to finite temperature baths. 

Our main results are twofold:
(i) We determine the phase diagram of the dissipative SSH model exposing analytically the impact of the electron-phonon coupling energy on the 
topologically-insulating and the trivial band-insulating phases.
The effective Hamiltonians reveal 
the renormalization of electron hoping terms due to the coupling to phonon baths, and non-trivially, 
the generation of phonon mediated electron-electron repulsion terms
within (between)  cells for intra (inter) cell coupling. This interaction term is shown to have a dramatic effect on the phase diagram, largely increasing the topological domain in the intercell coupling scheme. 
(ii) In the case of the  SSH model with an additional staggered potential, we show analytically that at strong coupling to the baths, this potential is suppressed with the model recovering the behavior of the unbiased SSH chain. 
At weak coupling to the baths, however, we show that a charge density wave phase develops in the staggered case. 

The paper is organized as follows. In Sec.~\ref{sec: Effective Hamiltonian approach}, we describe our analytic approach for obtaining an equilibrium state of a many-body system coupled to local bosonic baths at arbitrary coupling strength and at nonzero temperatures. In Sec.~\ref{sec: Hamiltonians studied}, we specify the dissipative SSH Hamiltonians under two different local coupling schemes: (i) Intracell coupling in Sec.~\ref{subsec: Intra-cell coupling} and (ii) Intercell coupling in Sec.~\ref{subsec: Inter-cell coupling}. We further analyze in Sec.~\ref{subsec: Staggered potential} how a staggered SSH model transforms under these two dissipative scenarios. In Sec.~\ref{sec: topological phase boundary}, we examine the topological phases of the these SSH chains at finite temperature using a topological marker suitable for density matrices, known as the \textit{ensemble geometric phase} (EGP). 
We summarize our findings and conclude in Sec.~\ref{sec: conclusion and discussion}.

\section{The RCPT approach}
\label{sec: Effective Hamiltonian approach}

We review here the \textit{reaction-coordinate polaron-transform} (RCPT) approach for obtaining the steady-state 
density matrix of a system in contact with bosonic environment(s) at arbitrary coupling strength. The method was developed in details and exemplified in Ref. \citenum{Anto_prx}. The RCPT technique generates an effective Hamiltonian, and it has been previously applied to impurity models and many-body spin systems \cite{Anto_prx,Anto_prb,Min_2024,Brenes_2024}.
It is worthwhile to note that previous studies applied the basic reaction coordinate mapping onto interacting fermionic quantum dot models \cite{Nazir18,GernotF18,GernotF19,GernotF21}. 
The construction of an effective Hamiltonian for a double quantum dot models was done in Ref. \citenum{Anto_prx}. 
%
However, here, for the first time, the RCPT method is applied to describe the structure of many-body fermionic lattices, specifically systems that exhibit a symmetry-protected topological phase. We explain the principles of the mapping in this section. In Sec. \ref{sec: Hamiltonians studied}, along with Appendix \ref{sec: Details} and \ref{sec: staggered potential}, we apply it onto the SSH model.
 
We begin by considering a generic open quantum system described by the following Hamiltonian:
\begin{equation}
\label{eq: starting Hamiltonian}
    \hat{H} = \hat{H}_S+\hat{S}\sum_kt_k\left(\hat{c}^\dagger_k+\hat{c}_k\right)+\sum_k\nu_k\hat{c}^\dagger_k\hat{c}_k.
\end{equation}
For simplicity, we explain the method using a single heat bath, but the mapping can be easily generalized to multiple baths. Here, $\hat{H}_S$ is an arbitrary many-body system's Hamiltonian and $\hat{S}$ is an operator acting on the system's Hilbert space that is coupled to the reservoir. The reservoir is modelled as a collection of harmonic oscillators, where $\hat{c}^\dagger_k~(\hat{c}_k)$ are the bosonic creation (annihilation) operators with frequency $\nu_k$ for the $k$-th harmonic mode. The system-bath coupling strength $t_k$ to the harmonic displacement operator is captured by the bath spectral density function, $J(\omega)=\sum_kt^2_k\delta(\omega-\nu_k)$.

The RCPT approach is a simple analytical mapping that involves two consecutive unitary transformations on the total Hamiltonian, Eq.~\eqref{eq: starting Hamiltonian}, followed by a controlled truncation. The aim of the mapping is to make the new system-bath coupling weaker than in the original model, while accurately imprinting the strong system-bath coupling energy into the  system Hamiltonian \cite{Anto_prx,Anto_prb,Min_2024,Brenes_2024}. The mapping generates an effective (eff), Hermitian Hamiltonian, reading $\hat{H}^\text{eff}=\hat{H}^\text{eff}_S(\lambda,\Omega)+\hat{H}^\text{eff}_I+\hat{H}^\text{eff}_B$, where the effective system Hamiltonian $\hat{H}^\text{eff}_S$ now explicitly depends on the original system-bath interaction energy scale ($\lambda$) and the characteristic bath frequency $(\Omega)$. Both of these parameters are obtained from the original bath spectral density function, $J(\omega)$. Let us now go through the mapping process step by step. 

The first of the two unitary transformations is the \textit{reaction coordinate} (RC) mapping \cite{Nazir18}.
Here, we extract a collective degree of freedom from the bosonic reservoir and incorporate it into the system. The RC is essentially a single harmonic oscillator mode with frequency $\Omega$, and it is coupled to the system via the coupling strength $\lambda$. As mentioned above, these parameters are defined based on the original spectral density function,  $ \lambda^2 = \frac{1}{\Omega}\int^\infty_0 \omega J(\omega)d\omega$ and $ \Omega^2  = \frac{\int^\infty_0 \omega^3 J(\omega)d\omega}{\int^\infty_0 \omega J(\omega)d\omega}$ \cite{Nazir18}. 

After the RC transform, the system does not directly couple to the residual bath, only indirectly, via the RC mode. Post RC mapping, the Hamiltonian reads
\begin{equation}
\begin{aligned}
    \hat{H}_\text{RC} = &\hat{H}_S+\lambda\hat{S}\left(\hat{a}^\dagger+\hat{a}\right)+\Omega\hat{a}^\dagger\hat{a}\\
    +&\left(\hat{a}^\dagger+\hat{a}\right)\sum_kf_k\left(\hat{b}^\dagger_k+\hat{b}_k\right)+\sum_k\omega_k\hat{b}^\dagger_k\hat{b}_k.
    \end{aligned}
\end{equation}
Here, $\hat{a}^\dagger~(\hat{a})$ is the creation (annihilation) operator for the RC mode and $\hat{b}^\dagger_k~(\hat{b}_k)$ are the creation (annihilation) operator for the residual bath modes. The RC unitary transformation is essentially a \textit{Bogoliubov} transformation that achieves $\lambda(\hat{a}+\hat{a}^\dagger)=\sum_kt_k(\hat{c}_k+\hat{c}^\dagger_k)$. The coupling between the RC and the residual bath is described by a different spectral density function, $J_\text{RC}(\omega)=\sum_kf^2_k\delta(\omega-\omega_k)$, where $f_k$ are the new coupling parameters. This new spectral density function can be found from the original spectral function \cite{Nazir18}. However, as we explain later in this Section, our ansatz for the equilibrium (long time) state of the system does not require knowledge of the new spectral function.
%
The RC transformation is constructed such that the scaling $\epsilon\times J(\omega)$  translate to $\lambda^2\propto \epsilon$ but
$J_{\text{RC}}(\omega)$ remains independent of $\epsilon$.

The second unitary transform is the \textit{polaron} transformation \cite{Cao_2016,Agarwal_2013}, given by $\hat{U}_P=\exp[\frac{\lambda}{\Omega}(\hat{a}^\dagger-\hat{a})\hat{S}]$, which is enacted on the system-RC Hilbert space while leaving the residual bath unaffected. This transform shifts the bosonic operator of the RC mode according to the state of the system (via the coupling to $\hat{S}$) as $\hat{a}^{(\dagger)}\rightarrow \hat{a}^{(\dagger)}-\frac{\lambda}{\Omega}\hat{S}$. The polaron transform 
seemingly complicates the Hamiltonian:
It imprints the RC coupling into the original system thereby weakening the coupling between the RC and the system. However, in addition, it generates a new coupling term between the original system and the residual bath. It may also generate bath-induced interactions between systems' degrees of freedom. Post polaron transform, the Hamiltonian reads
\bea
    \hat{H}_\text{RC-P} &=& 
    \hat{U}_P\hat{H}_\text{RC}\hat{U}^\dagger_P 
    \nonumber\\
    &=&
    \hat{U}_P\hat{H}_S\hat{U}^\dagger_P-\frac{\lambda^2}{\Omega}\hat{S}^2+\Omega\hat{a}^\dagger\hat{a}
    \nonumber\\
    &+&\left(\hat{a}^\dagger+\hat{a}-\frac{2\lambda}{\Omega}\hat{S}\right)\sum_kf_k\left(\hat{b}^\dagger_k+\hat{b}_k\right)+\sum_k\omega_k\hat{b}^\dagger_k\hat{b}_k.
    \nonumber\\
\eea
The bath-induced interaction term is given by $-\frac{\lambda^2}{\Omega}\hat{S}^2$.

After the polaron transformation, the Hamiltonian is truncated by projecting it onto the ground state of the polaron-transformed RC. Such a restriction to a sub-manifold of the Hamiltonian is justified under the assumption that $\Omega$, the characteristic frequency of the bath, is the largest energy scale in the problem. 
This truncation eliminates terms in $\hat{H}_\text{RC-P}$, those that are proportional to $\hat{a}^\dagger\hat{a}$, as well as $\hat{a}^\dagger+\hat{a}$. Hence, the final effective Hamiltonian after the RCPT machinery is given by
\begin{equation}
\label{eq: effective total Hamiltonian}
    \hat{H}^\text{eff} = \hat{H}^\text{eff}_S(\lambda,\Omega)-\hat{S}\sum_k\frac{2\lambda f_k}{\Omega}\left(\hat{b}^\dagger_k+\hat{b}_k\right)+\omega_k\hat{b}^\dagger_k\hat{b}_k,
\end{equation}
where 
\bea 
\hat{H}^\text{eff}_S(\lambda,\Omega) = \bra{0}\hat{U}_P\hat{H}_S\hat{U}^\dagger_P\ket{0}-\frac{\lambda^2}{\Omega}\hat{S}^2,
\label{eq:HSeff}
\eea
with $\ket{0}$ being the ground state of the RC mode. 

The elegance of the RCPT treatment is that in Eq. (\ref{eq: effective total Hamiltonian}), the system Hamiltonian is weakly coupled to the residual bath (explained next), but the total Hamiltonian structure is conserved in comparison to Eq.~\eqref{eq: starting Hamiltonian}. 

Let us discuss some key features of Eq.~\eqref{eq: effective total Hamiltonian}. 

(1) The coupling of the effective system to the residual bath is determined by the set of coupling energies $\{f_k\}$, distinct from the original coupling energies \{$t_k$\}. The RC transformation is useful when the residual coupling is weak, achieved through the design of the transformation.
The new coupling parameters  are furthermore re-scaled according to  $f_k\rightarrow 2\lambda f_k/\Omega$. With this residual coupling strength acting as a perturbative parameter, strong system-bath coupling effects can be examined as a function of $\lambda$, the system-bath interaction energy scale as defined in the starting point model. After the polaron transformation, the bath spectral density of the effective bath is further dressed by the bath parameters,$J^\text{eff}(\omega)=\frac{4\lambda^2}{\Omega^2}J_\text{RC}(\omega)$. 

(2) In principle, the effective spectral density can be obtained from any original spectral density function, $J(\omega)$. However, the RCPT method is useful (e.g., it creates a good approximation for the equilibrium state of the system) if the transformed spectral density function corresponds to a weak residual coupling. For example, a Brownian spectral density function that describes the original system-bath coupling,
 $   J(\omega) = \frac{4\gamma \Omega^2\lambda^2\omega}{(\omega^2-\Omega^2)^2+(2\pi\gamma \Omega\omega)^2}$,
is peaked at energy $\Omega$ with width parameter $\gamma$. Post RC mapping, this spectral density becomes Ohmic, given by $J^\text{RC}(\omega)\propto \gamma\omega$. The parameter that described the width of the spectral function in the original model now becomes a prefactor that dictates the residual coupling. If we work in the limit of $\gamma \ll 1$, we are able to study strong system-bath coupling physics (in the original model) in a computationally economic manner (in the effective model).


Given $\hat{H}^\text{eff}_S$,  we make the ansatz that the approximate equilibrium state of the system is the canonical Gibbs state \cite{Cresser_2021}, however written here with respect to the effective system Hamiltonian, $\hat{\rho}=e^{-\beta\hat{H}^\text{eff}_S}/\Tr(e^{-\beta\hat{H}^\text{eff}_S})$. Here, $\beta$ is the inverse temperature of the bath. An equilibrium observable, $\hat{\mathcal{O}}$, can then be computed as $\langle\hat{\mathcal{O}}\rangle=\Tr(\hat{\rho}\hat{\mathcal{O}})$.

Since we are only interested in the equilibrium state of the system, we do not need to specify the spectral function of the residual bath, 
as it only dictates the rate to reach equilibrium, rather than the state itself. 
Moreover, we do not need to commit to a particular spectral function for the original model; we only need to know the values of the parameters $\lambda$ and $\Omega$, characterizing the system-bath interaction energy and the bath characteristic frequency, respectively.
Our conclusions here, as to the topological phase diagram of the dissipative SSH model, hold for variety of spectral functions, with the only constrain that the residual coupling is weak. An example for a valid choice is a narrow Brownian function, as mentioned above.


\section{The dissipative SSH model with different coupling schemes}
\label{sec: Hamiltonians studied}

In this section, we introduce the SSH Hamiltonian under two different dissipation scenarios, depicted in Fig.~\ref{fig:figure 1}.
The model describes spinless fermions, which are hopping on a one-dimensional chain with staggered hopping tunneling elements. The chain includes $L$ cells, marked by the sublattice indices $A$ and $B$. We begin by describing the model with uniform-equal site energies. We later introduce the Rice-Mele model in which the A and B sites are distinguished by their site energy.

The hopping Hamiltonian of the SSH model is given by
\bea
\hat{H}_\text{SSH}(v,w) =&-&v\sum^L_{i=1}\left(\hat{d}^\dagger_{i,A}\hat{d}_{i,B}+h.c.\right)
\nonumber\\
&-&w\sum^{L-1}_{i=1}\left(\hat{d}^\dagger_{i,B}\hat{d}_{i+1,A}+h.c.\right).
\label{eq:HSSH}
\eea
Here, $v$ and $w$ are the alternating hopping amplitudes, corresponding to the intrrcell ($v$) and intercell ($w$) couplings.
The operator $\hat{d}^\dagger_{i,a}~(\hat{d}_{i,a})$ is the spinless fermion creation~(annihilation) operator on the $i$-th unit cell with a sublattice index $a\in\{A,B\}$. 

The closed-system zero $T$ SSH model at half-filling is the simplest example of a symmetry-protected topological insulator \cite{BookTI}. The alternating hopping amplitudes give rise to the chiral (sublattice) symmetry,
and the chain has two topologically distinct phases:  The region $v<w$  corresponds to the topological insulator (TI) phase; $v>w$ is the trivial band insulator (BI) phase. 

Considering now the open-system model, we couple the SSH chain to a set of local heat baths at each unit cell, or at each unit cell boundary, depending on the coupling scheme.
The intracell dissipation scheme is depicted in Fig. \ref{fig:figure 1}(a), while the intercell scheme is depicted in \ref{fig:figure 1}(b).
In the following subsections, we write down the total Hamiltonian for each of these dissipative scenarios. We then present the   effective SSH Hamiltonians of the system, which serve to study the equilibrium state of the models at arbitrary system-bath couplings and nonzero temperature.
We also report on the effect of the two coupling schemes upon introducing an on-site staggered potential to the SSH model.

\subsection{Intracell electron-phonon coupling scheme} 
\label{subsec: Intra-cell coupling}

We consider the SSH model (\ref{eq:HSSH}) as the system Hamiltonian, and add an intracell dissipative coupling, depicted in Fig.~\ref{fig:figure 1} (a). 
The total Hamiltonian is given by
\begin{equation}
\begin{aligned}
\hat{H} = \hat{H}_{\text{SSH}}(v,w)+\hat{H}_B+\hat{H}_I.
    \end{aligned}
\end{equation}
The bath Hamiltonian includes the collection of independent harmonic baths,
\bea
\hat{H}_B&=&\sum_{n,k}\nu_{k}\hat{c}^\dagger_{n,k}\hat{c}_{n,k}.
\eea
The system-bath interaction coupled the displacements of bosons to electron hopping between sites,
\bea
\hat{H}_I=
\sum^L_n\hat{S}^\text{intra}_n\sum_kt_{k}(\hat{c}^\dagger_{n,k}+\hat{c}_{n,k}), 
\eea
where
\bea\hat{S}^\text{intra}_n=\hat{d}^\dagger_{n,A}\hat{d}_{n,B}+h.c..
\eea
For simplicity, we assume that all the baths have the same set of frequencies $\nu_k$ and coupling energies to the system, $t_k$. 
However, calculations can be readily generalized beyond that. Here, $n$ denotes the bath index, which we interchangeably employ to identify the unit cell, $i$ (for the intracell coupling scheme, they are indeed identical). From here on, the SSH chain is understood to have a periodic boundary condition. 

The steps involved in the RCPT mapping of this model are presented in Appendix.~\ref{sec: Details}. It yields the following effective system Hamiltonian, as formulated in Eq. (\ref{eq:HSeff}), 
\begin{equation}
\label{eq: intra EFFH}
\begin{aligned}
\hat{H}^\text{intra}_\text{SSH,eff} =& \hat{H}_\text{SSH}(v,\tilde{w})+\hat{H}^\text{intra}_\text{MB}\\
=& \hat{H}_\text{SSH}(v,\tilde{w})-\frac{\lambda^2}{\Omega}\sum^L_{i=1}\left(\hat{n}_{i,A}-\hat{n}_{i,B}\right)^2
\end{aligned}
\end{equation}
Here, $\tilde{w}=we^{-\lambda^2/\Omega^2}$ and $\hat{n}_{i,a}=\hat{d}^\dagger_{i,a}\hat{d}_{i,a}$. 
The RCPT machinery reveals the effect of the intracell coupling through two aspects:
(i) The {\it intercell} hopping amplitude $w$ is exponentially suppressed by the electron-phonon coupling energy.  
(ii) The baths generate a many-body (MB) interaction term between spinless fermions on sites $A$ and $B$, see the last term of Eq.~\eqref{eq: intra EFFH}. We denote this MB interaction by $\hat{H}^\text{intra}_\text{MB}$. 

The suppression of $w$ already hints on the detrimental behavior of the intracell coupling scheme on the topologically insulating phase, since the topological criterion for the bare (nondissipative) SSH model is that $v<w$. That is, even if one starts with a topological phase, sufficiently strong coupling with local baths would lead to a trivial band insulator. Not less importantly, the MB interaction term adds an energy cost of $2\lambda^2/\Omega$ if the unit cell is doubly occupied. Intuitively, such a term can be rationalized as the coupling scheme favors $v$ hopping, which is  possible if the unit cell only is singly occupied. As we argue in Sec.~\ref{sec: topological phase boundary}, it is actually this bath-induced MB interaction term that has a more detrimental effect on the topological phase than the direct suppression of $w$ through its dressing. It was recently reported that such dimerized interaction has a significant impact on the topological phase of the SSH model \cite{Mondal_2022}. 


\subsection{Intercell electron-phonon coupling scheme}
\label{subsec: Inter-cell coupling}

We consider next the intercell coupling scheme,  depicted in Fig.~\ref{fig:figure 1} (b). The model is described by the  total Hamiltonian 
\begin{equation}
    \hat{H}=\hat{H}_\text{SSH}(v,w)+\hat{H}_B+\hat{H}_I.
\end{equation}
The interaction electron-boson baths Hamiltonian,
\bea
\hat{H}_I=\sum^{L-1}_n\hat{S}^\text{inter}_n\sum_kt_k\left(\hat{c}^\dagger_{n,k}+\hat{c}_{n,k}\right),
\eea
now involves a different operator,
\bea 
\hat{S}^\text{inter}_n=\hat{d}^\dagger_{n+1,A}\hat{d}_{n,B}+h.c.,
\eea
with the collection of baths given by
\bea
\hat{H}_B&=&\sum_{n,k}\nu_k\hat{c}^\dagger_{n,k}\hat{c}_{n,k}.
\eea
Again, a periodic boundary condition is implied. 

By symmetry, see Appendix.~\ref{sec: Details} for details, the effective system Hamiltonian after the RCPT machinery is given by
\begin{equation}
\label{eq: inter EFFH}
\begin{aligned}\hat{H}^\text{inter}_\text{SSH,eff} = &\hat{H}_\text{SSH}(\tilde{v},w)+\hat{H}^\text{inter}_\text{MB}\\
&\hat{H}_\text{SSH}(\tilde{v},w)-\frac{\lambda^2}{\Omega}\sum^{L-1}_{i=1}\left(\hat{n}_{i+1,A}-\hat{n}_{i,B}\right)^2,
\end{aligned}
\end{equation}
with the dressing of the intracell tunneling according to $\tilde{v}=ve^{-\lambda^2/\Omega^2}$.

The effective system Hamiltonian for the intercell coupling, Eq. (\ref{eq: inter EFFH}), is analogous to that of the intracell coupling, Eq. (\ref{eq: intra EFFH}). The distinctions being that in the intercell model, the {\it intracell} $v$ hopping is exponentially suppressed, instead of the $w$ hopping, and that the fermion-fermion interaction terms are built up at the unit-cell boundaries, rather than within the unit cell. Due to the suppression of the intracell $v$ hopping, (while the $w$ hopping is retained), we  infer that the intercell coupling scheme is advantageous to the topological phase. We again emphasize that we later show that the bath-induced MB term $\hat{H}^\text{inter}_\text{MB}$ brings about the most dramatic control of the topological phase.

\subsection{The Rice-Mele model: An SSH model with a staggered potential}
\label{subsec: Staggered potential}

The Rice-Mele (RM) model generalizes the SSH model, including
 an on-site staggered potential. That is,  on top of the usual SSH chain, Eq. (\ref{eq:HSSH}), the following term is added,
\begin{equation}
    \hat{H}_\text{SP}(u) = u\sum^L_{i=1}\left(\hat{d}^\dagger_{i,A}\hat{d}_{i,A}-\hat{d}^\dagger_{i,B}\hat{d}_{i,B}\right).
\end{equation}
Here, $u$ is the magnitude of the on-site potential. 
This term breaks the sublattice symmetry, and thereby it allows a smooth connection between the topologically distinct phases of the SSH model, without closing the gap. 
The RM system's Hamiltonian, prior to the RCPT transform, reads 
$\hat{H}_\text{SSH}(v,w)+\hat{H}_\text{SP}(u)$. After the RCPT  procedure, the staggered-potential term simply changes to
\begin{equation}
    \hat{H}_\text{SP}(u)\rightarrow \hat{H}_\text{SP}(\tilde{u}),
    \label{eq:RM}
\end{equation}
where $\tilde{u} = ue^{-\frac{2\lambda^2}{\Omega^2}}$, regardless of the dissipative coupling scheme (intracell or intercell). For details, see  Appendix.~\ref{sec: staggered potential}. 
The effect of the local bosonic baths is therefore to suppress the on-site energy imbalance: 
the onsite potentials of the $A$ and $B$ sites approach their average value. A more general choice of on-site potentials is discussed in Appendix~\ref{sec: staggered potential}. 

\section{Results: Dissipation control of the topological phase boundary}
\label{sec: topological phase boundary}

We describe here our results on the phase diagram of the SSH model with either intracell or intercell electron-phonon couplings, as well as the Rice-Mele model variant.

Given the effective system Hamiltonians obtained via the RCPT machinery for the two coupling schemes, Eq.~\eqref{eq: intra EFFH} and Eq.~\eqref{eq: inter EFFH}, our goal now is to study 
how the the boundary between the topological and normal phases is modified by the coupling to local baths. We emphasize that the obtained effective Hamiltonians correspond to the SSH chain when subjected to a non-perturbative system-bath interaction. 

By inspecting Eqs.~\eqref{eq: intra EFFH} and Eq.~\eqref{eq: inter EFFH}, we have already gathered that the intracell coupling scheme should be detrimental to the topological phase, while the intercell coupling scheme should be advantageous. In this section, we quantify this claim. 
First, we present below a well-studied topological marker that characterizes the two topologically-distinct phases of the SSH chain. 
This measure is suitable for systems described by density matrices in a mixed state. 
Next, we present numerical results: We compute this topological invariant through an exact diagonalization (ED) technique.


Due to the assumption of a nonzero temperature in the construction of the equilibrium state, standard definitions for the topological invariant, such as the Zak phase \cite{Zak_1989}, do not apply, since they are only useful to characterize the ground-state.
We therefore adopt the mixed state generalization of the \textit{Resta's polarization} measure \cite{Resta_1998}, which is directly related to the topological invariant for density matrices \cite{Bardyn_2018,Wawer_2021_2}. This so called \textit{ensemble geometric phase} (EGP) is given by
\bea
    \phi_E &=&  \Im \ln \Tr(\hat{\rho}\hat{T}) 
    \nonumber\\
    &=& \Im \ln \Tr(\hat{\rho}e^{\frac{2\pi i}{L}\hat{X}}).
    \label{eq:EGPd}
\eea
Here, $\hat{\rho}$ is the density matrix of the system and $\hat{X}=\sum_n [n\hat{d}^\dagger_{n,A}\hat{d}_{n,A}+(n+1/2)\hat{d}^\dagger_{n,B}\hat{d}_{n,B}]$ is the many-particle center of mass operator, defined for the chain model. This topological marker faithfully captures the geometric phase for mixed quantum states. To compute the EGP, we numerically diagonalize the effective system Hamiltonian, given either by Eq.~\eqref{eq: intra EFFH} or Eq.~\eqref{eq: inter EFFH}. We assume half-filling and adopt a standard ED procedure. Based on the eigenvectors and eigenvalues of the many-body Hamiltonian, we compute the thermal average of $\hat{T}$, see Eq. (\ref{eq:EGPd}). The phase of this thermal average corresponds to the EGP. 

Perhaps, the only caveat of the EGP is that it does not capture the critical temperature $T_c$; when $T>T_c$, the topological phase ceases to exist. That is, if the temperature is too high, comparable to the gap, 
we expect the edge modes, protected by the gap, to vanish. There exists another topological marker for density matrices, known as the \textit{Ulhmann phase} (UP). This measure is a formal generalization of the Zak phase \cite{Viyuela_2014}. Unlike the EGP, the UP does predict a critical temperature $T_c$ above which the topological phase vanishes. For our simulations, we choose temperatures that are lower than the critical temperature:
The UP predicts the critical temperature to be 
$T_c=1/\ln(2+\sqrt{3})$ \cite{Viyuela_2014}, 
which is about 38\% of the constant energy gap in the fully dimerized limit. However, the reliability of the UP as a faithful topological invariant is in doubt due to the insufficient stringency of the gauge structure, which always leads to a topologically trivial case \cite{Budich_2015,Molignini_2023}. Furthermore, extending such a measure to higher dimensions cannot be uniquely defined by the UP \cite{Budich_2015,Molignini_2023}. For this reason, we employ the EGP rather than the UP, as it naturally extends the familiar connection between the geometrical phase and polarization to finite temperatures. 

\begin{figure}[htpb]
\fontsize{6}{10}\selectfont 
\centering
\includegraphics[width=0.9\columnwidth]{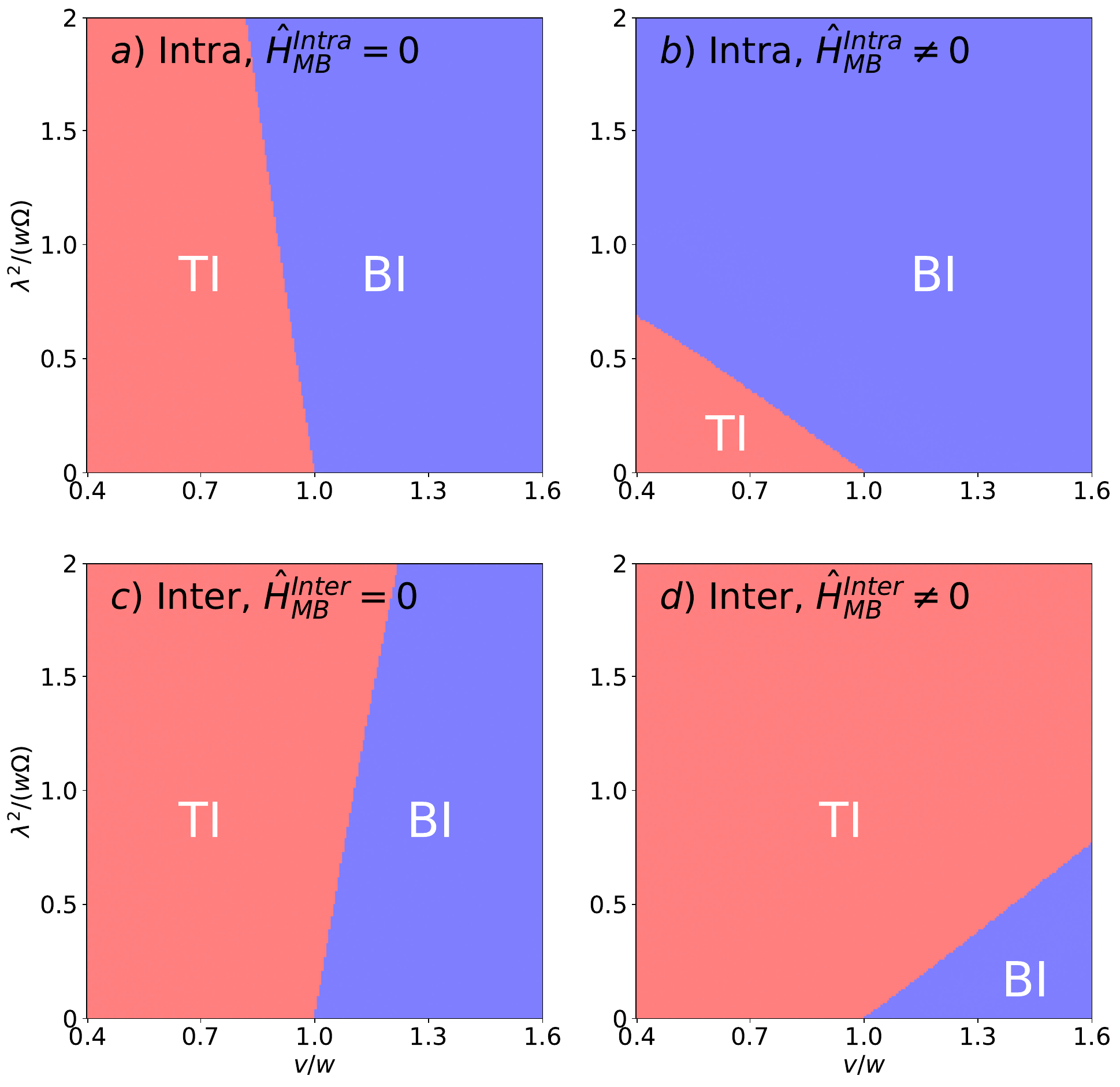}
\caption{Phase diagram of the SSH model.
The phases are characterized by the ensemble geometric phase. The red phase corresponds to an EGP of $\pi/2$ (Topological Insulator, TI) while the blue phase corresponds to an EGP of $-\pi/2$ (Band Insulator, BI).
We study the intracell (a)-(b) and intercell (c)-(d) coupling schemes, without (left) or with (right) the bath-induced many-body interaction term, $\hat{H}_\text{MB}$. 
 (a), (c): Without the MB term, the tilting of the phase boundary towards left(right) is solely attributed to the suppression of the hopping parameter $w(v)$ for the intra(inter)-cell coupling scheme.  (b), (d): Taking into account the bath-induced many-body interaction term drastically diminishes or expands the topological phase. The chain includes $L=6$ cells (12 sites). We fix the intercell tunneling $w=1$ and vary $v$. Other parameters are $T=0.5$, baths' characteristic frequency $\Omega=10$, and $u=0$, corresponding to an unbiased chain.}
\label{fig:figure 2}
\end{figure}

\begin{figure}[htpb]
\fontsize{6}{10}\selectfont 
\vspace{-3mm}
\centering
\includegraphics[width=0.9\columnwidth]{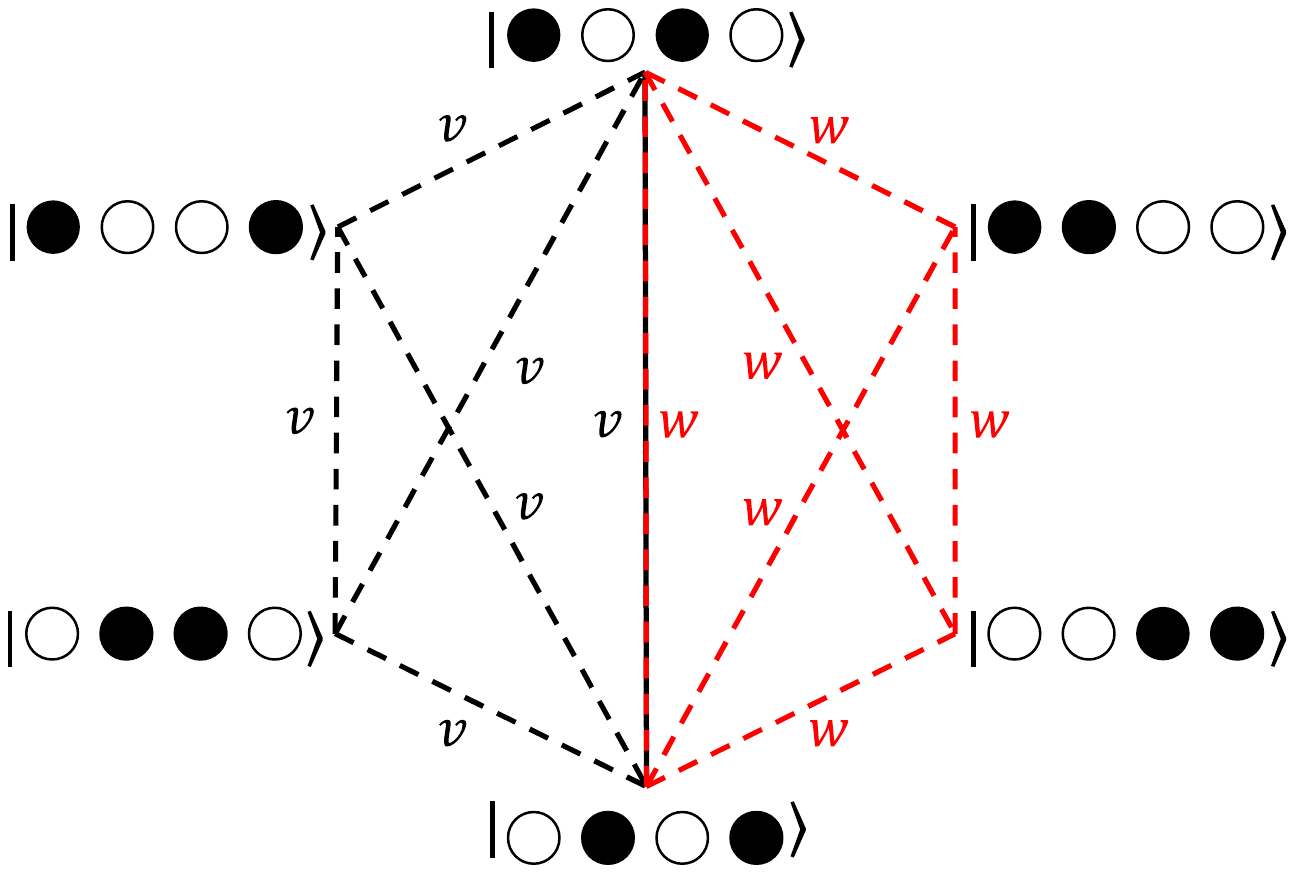}
\caption{Graph depiction of the six basis states of an $L=2$ cell system at half-filling, showing the states connectivity via hopping amplitudes. 
In the intracell coupling scheme, the two basis states on the right side $\ket{\bullet\bullet\circ\circ}$ and $\ket{\circ\circ\bullet\bullet}$ are energetically costly.  As such, it is less likely for electrons to jump to these state with increasing system-bath coupling strength, as it increases this energetic cost. This scenario effectively weakens the $w$ hopping, leading to the suppression of the topological phase. 
Conversely, in the intercell coupling scheme, the two basis states on the left edge,  $\ket{\bullet\circ\circ\bullet}$ and $\ket{\circ\bullet\bullet \circ}$  are energetically unfavorable to occupy. This effectively suppresses the $v$ hopping energy, thereby favoring the topological phase.}
\label{fig:figure 3}
\end{figure}

\subsection{Dissipative SSH model: Boundary of the topological phase} 

In Fig.~\ref{fig:figure 2}, we plot the phase diagram of the dissipative SSH chain under the two coupling schemes, as characterized by the EGP. 
We clearly note that an intracell coupling scheme is detrimental to the topological phase, while the intercell coupling scheme is advantageous, in accordance with previous simulations \cite{Pavan_2024}. 
However, with our RCPT approach, we are able to uncover the microscopic factors behind the expansion or suppression of the topological phase. That is, we find that it is the bath-induced many-body interaction term that drastically alters the phase boundary, more than the renormalization of tunneling elements.
To demonstrate that, we first turn off the MB interaction terms, and study the phase diagram while only maintaining the bath-induced dressing of parameters. Results are presented in Fig.~\ref{fig:figure 2}(a) and (c). In this case, the phase boundary is simply dictated by $v<\tilde{w}(\tilde{v}<w)$ for the intracell (intercell) coupling. 
Although the suppression of the hopping parameters indeed hampers (enhances) the parameter space for the topological phase,  in Fig.~\ref{fig:figure 2}(b) and (d) we show that it is the bath-induced many-body interaction term that more drastically impacts the phase boundary. 

It is worth noting that the MB interaction terms must be taken into account explicitly. A mean-field approximation to the interacting terms will lead to an incorrect phase diagram that completely misses the correct reshaping of the phase boundary. We detail on this issue in Appendix.~\ref{sec: MF theory}.

How do we explain the impact of the bath-induced MB interaction terms on the boundary of the topological phase?  
We argue that the MB interaction effectively suppresses the appropriate hopping amplitudes---depending on the coupling scheme. To substantiate and provide 
an intuitive picture
we display in Fig.~\ref{fig:figure 3} a graph depiction of the basis states for an $L=2$ chain at half-filling. 
A two cell chain with two electrons is represented by six basis states. We label the states and mark how they are connected to one another via the hopping terms, $v$ and $w$.

Let us first consider the intracell coupling scheme.
Recall that $H_{\text{MB}}^{\text{intra}}=
-\frac{\lambda^2}{\Omega}
\sum^L_{i=1}\left( \hat{n}_{i,A}-\hat{n}_{i,B}\right)^2$  
according to Eq. (\ref{eq: intra EFFH}).
As such, the two states depicted at the rightmost edge of the hexagon are unfavourable to be occupied. They each require an energy cost of $2\lambda^2/\Omega$ once interaction terms are turned on. 
This makes it energetically unfavorable to ``visit" these states, thereby effectively suppressing the $w$ intracell hopping. This effective suppression is apparently the dominant mechanism responsible for adjusting the phase boundary due to coupling to the baths, as shown in Fig.~\ref{fig:figure 2}(b), compared to (a). Altogether, in the intracell coupling scenario, the explicit suppression of $w\to\tilde w$, combined with the MB term, leading to an additional effective reduction of this hoping term, result in a strong suppression of the topological region with increasing electron-boson coupling. 

On the other hand, in the intercell scheme the MB term involves two cells,  $H_{\text{MB}}^{\text{inter}}=
-\frac{\lambda^2}{\Omega}
\sum^L_{i=1}\left( \hat{n}_{i+1,A}-\hat{n}_{i,B}\right)^2$.  
Then, it is the two states on the leftmost edge that require an energy cost to be occupied. 
In this case, the intracell hopping $v$ is effectively suppressed, leading to the expansion of the window of the topological phase compared to the zero-coupling case, compare Fig.~\ref{fig:figure 2}(c) to (d).

\begin{figure}[h]
\fontsize{6}{10}\selectfont 
\centering
\includegraphics[width=1.0\columnwidth]{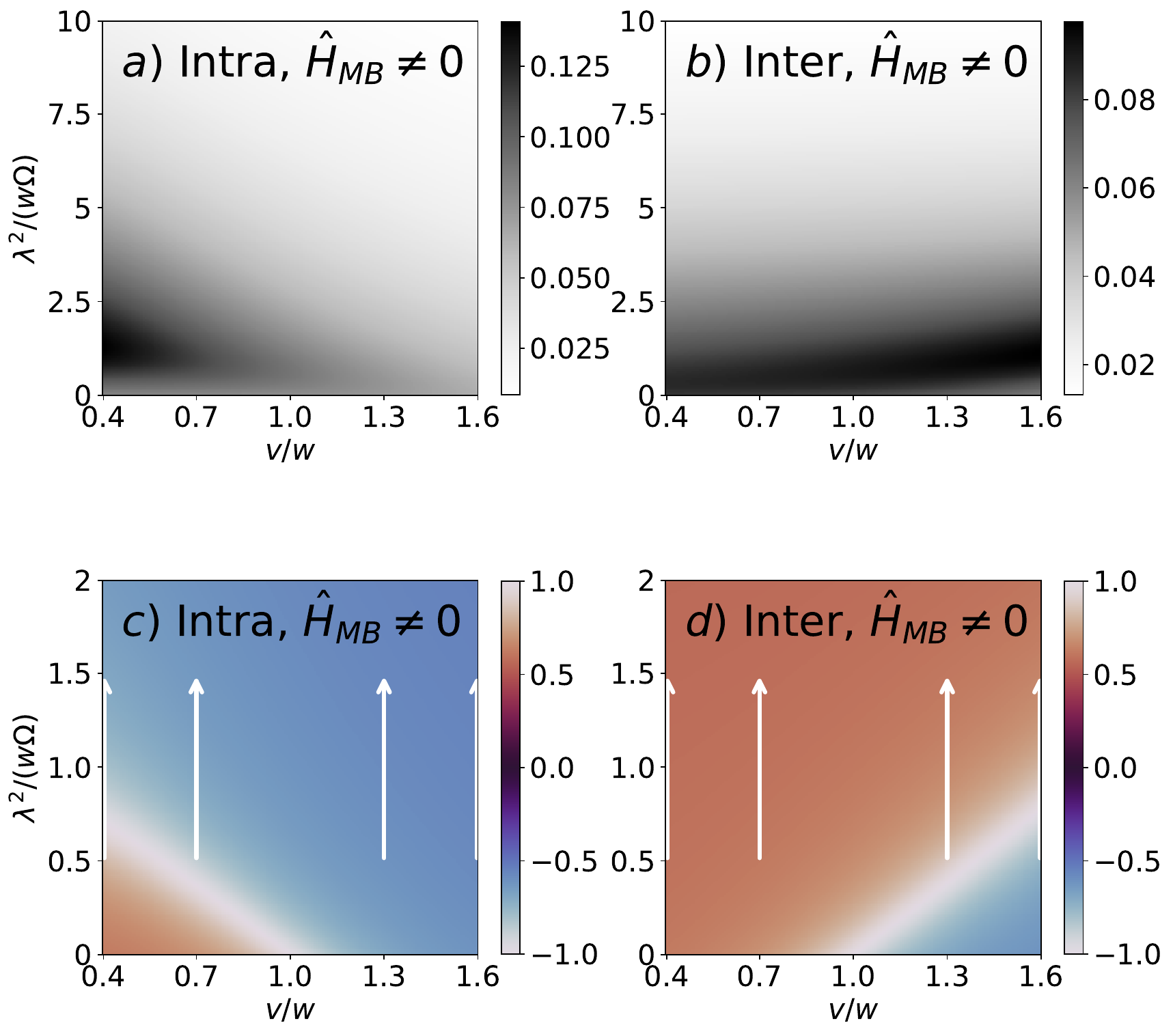}
\caption{Study of the Rice-Mele model, an SSH model with a staggared potential.
(a)-(b) The CDW order parameter $\mathcal{X}$ for (a) intracell and (b) intercell coupling schemes. As the coupling strength increases, the CDW order parameter is suppressed.
(c)-(d)  The EGP (in multiple of $\pi$) 
for the Rice-Mele chain.  When the staggered potential is introduced to the SSH chain, it loses its quantized EGP of \(\pi/2\) and \(-\pi/2\) for the topological and trivial phases, respectively. White arrows indicate increasing coupling strength.
Parameters are $L=6$ , a staggered potential $u=0.2$, $w=1$, $T=0.5$, and $\Omega=10$.
}
\label{fig:figure 4}
\end{figure}

\begin{figure}[htpb]
\fontsize{6}{10}\selectfont 
\centering
\includegraphics[width=0.9\columnwidth]{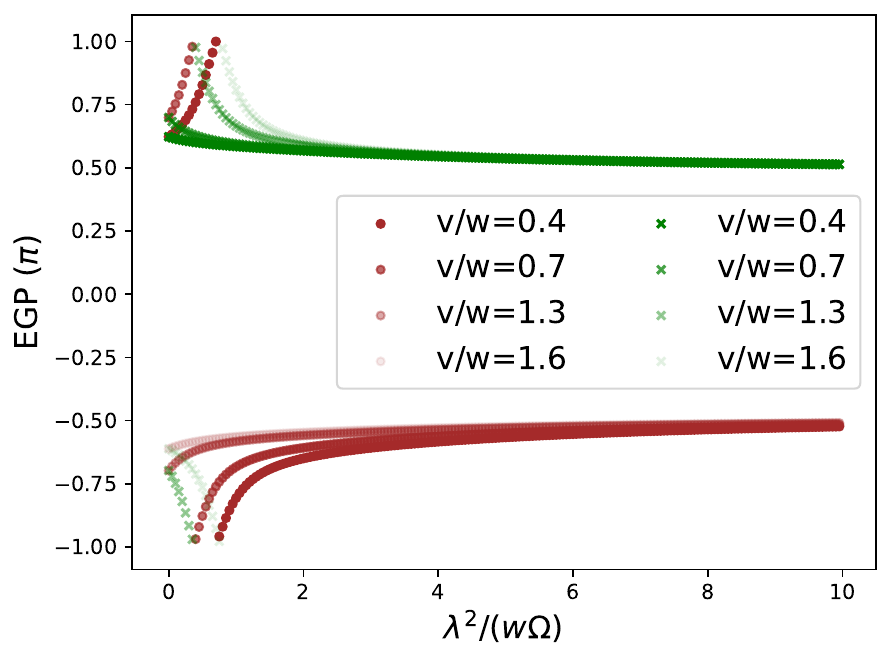}
\caption{Analysis of the Rice-Mele model showing the EGP for an intracell (brown circle) and intercell (green cross) coupling schemes. Parameters are identical to that in Fig.~\ref{fig:figure 4}. 
As the coupling strength increases, the intracell scheme approaches an EGP value of $-\pi/2$, while the intercell scheme goes to $\pi/2$. That is, the sublattice symmetry is restored 
with increasing $\lambda$,
and the EGP approaches the values for the SSH model.}
\label{fig:figure 5}
\end{figure}
\subsection{The dissipative Rice-Mele model}

As an additional variant of the SSH model,
we study here the effect of an on-site staggered potential on the topological phase boundary. 
We define the charge density wave (CDW) order parameter as 
$\mathcal{X}=\frac{1}{L}\sum^L_{i=1}\langle\hat{n}_{i,A}-\hat{n}_{i,B}\rangle$,
and present it in Fig.~\ref{fig:figure 4}
for (a) intracell and (b) intercell coupling schemes. Here, the $\langle\dots\rangle$ is understood as a thermal average with respect to the 
thermal state.
The staggered potential is taken to be $u=0.2$. The rest of the parameters are taken identical to those in Fig.~\ref{fig:figure 2}. 

We find that as the electron-phonon coupling energy is increased, $\mathcal{\chi}$ approaches zero. This can be explained by the exponential suppression of the staggered potential, as exposed in the effective Hamiltonian, Eq. (\ref{eq:RM}).
As such, at strong coupling to the baths, the EGP value that the RM model is expected to approach is that of the SSH model, depending on the coupling scheme. This is shown in Fig.~\ref{fig:figure 4} (c)-(d): At weak coupling,
the EGP deviates from quantized values, $\pm\pi/2$.
However, as the coupling energy to the baths is increased, the EGP approached $-\pi/2$ $(+\pi/2)$ for the intracell (intercell) coupling scheme (follow the direction of white arrows). Although the sign of the EGP is maintained in comparison to the phase diagram depicted in Fig.~\ref{fig:figure 2}, there is no topological phase transition presented. Instead, we obverse only an asymptotic approach to the quantized EGP values. 

Complementing the map of the EGP, in Fig.~\ref{fig:figure 5}, we plot the EGP for sample values of $v/w$ ratios. 
These are the EGP values along the eight white arrows shown in Fig.~\ref{fig:figure 4} (c) and (d). 
We plot four curves (green crosses) corresponding to the intercell electron-phonon coupling scheme and four plots (brown circles) for the intracell scheme. As one increases $\lambda$, the EGP asymptotically approaches $\pi/2$ ($-\pi/2$) for the intercell (intracell) coupling schemes, as expected.

\begin{figure}[htpb]
\vspace{-7mm}
\fontsize{6}{10}\selectfont 
\centering
\includegraphics[width=1.0\columnwidth]{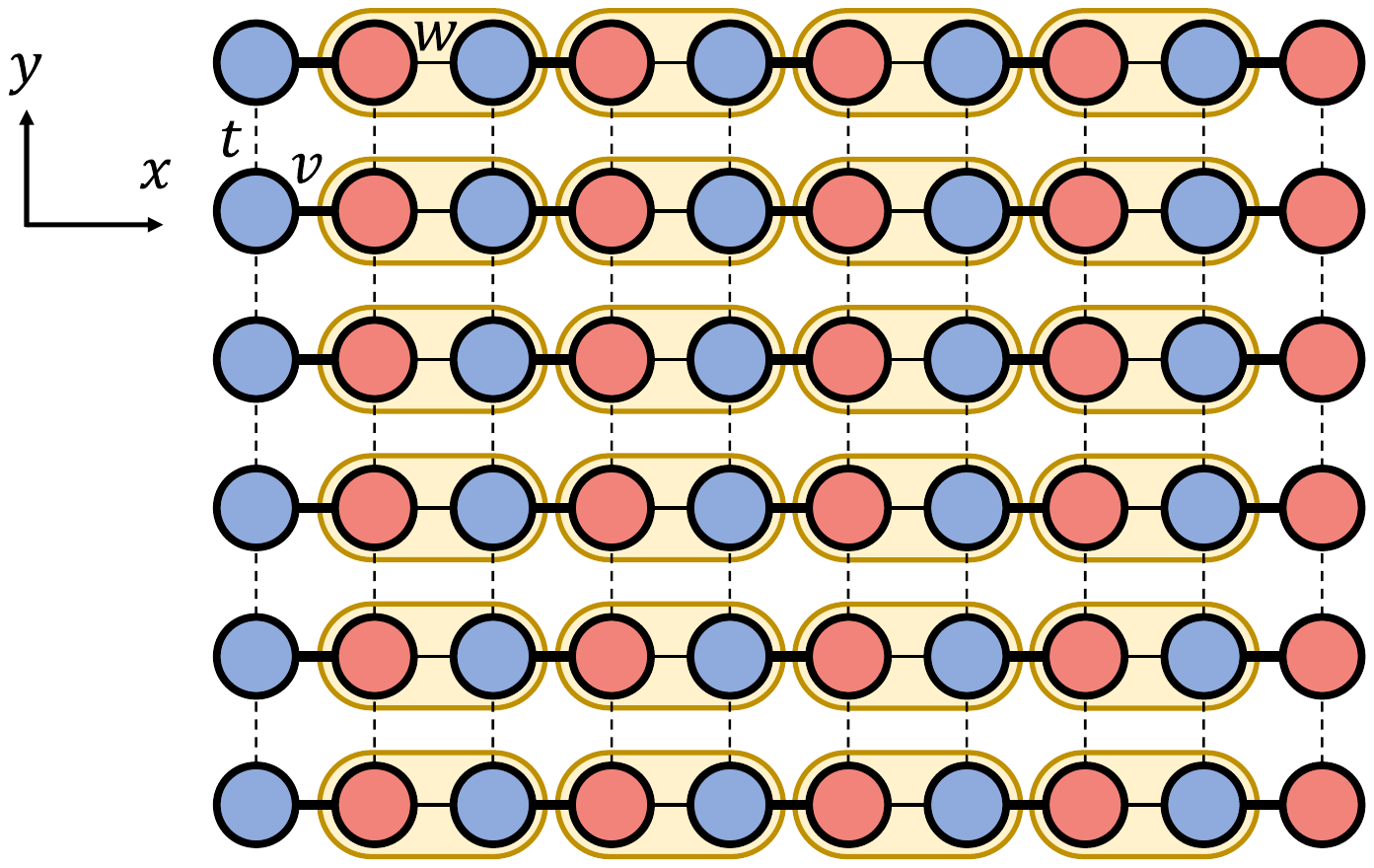}
\vspace{-7mm}
\caption{Two dimensional generalization of the SSH chain with intercell coupling to heat baths. This coupling scheme will suppress $v$ and the bulk $t$ hopping, but retain the boundary $t$ hopping amplitudes.}
\label{fig:figure 6}
\end{figure}

\section{conclusion and discussion}
\label{sec: conclusion and discussion}

In this paper, using a semi analytical tool, we studied how the boundary between the topologically insulating phase and the trivially insulating phase of the SSH chain was modified by coupling the system to local heat baths. We adopted either  an intracell or intercell coupling schemes. In both coupling models, the construction of an effective system Hamiltonian using the RCPT approach uncovered the underlying physics leading to the modification of the phase diagram due to the coupling to baths:
The effective Hamiltonian revealed the renormalization of electron hoping terms due to the coupling to thermal baths, and the generation of bath mediated electron-electron interaction (repulsion) terms within (between) cells for intra (inter) cell coupling.  To analyze the role of the thermal environments in modifying the topological insulator-band insulation boundary,
we computed the ensemble geometric phase, 
a topological invariant for the density matrices. 

The effective Hamiltonian approach allowed as to identify the dominant factor behind the baths' effect on the topological-normal insulator phase boundary by toggling the bath-induced fermion-fermion interaction terms and examining their effect on the phase boundary. It turned out that, compared to the direct renormalization of tunneling elements, the bath-mediated dimerized interaction had a  greater impact on the fate of the topological phase.
Overall, our study clarifies the origin of the observation that whereas intracell electron-bath interaction is detrimental for topology, an intercell coupling favors the development of the topological phase \cite{Pavan_2024}.

We also studied the Rice-Mele model, an SSH chain with an on-site staggered potential. We found that these potentials are exponentially suppressed in both coupling schemes when increasing the couplings to the baths, thereby bringing the deviated EGP values back to their quantized values of $\pm \pi/2$, and approaching the behavior of the non staggered dissipative SSH model. 

Our mapping approach is not limited to a specific spectral density function of the bath, but it only requires the residual coupling (after the Markovian embedding) to be weak. As such, our analysis suggests that the enhancement (suppression) of the topological window in the intercell (intracell) coupling scheme is a general phenomenon, which is not limited to a specific type of structured baths.  



We now comment on future applications of this study. As the RCPT method can be easily generalized to higher dimensions, perhaps the most immediate next problem in the examination of dissipative phases could be the two-dimensional (2D) SSH model. 
In Fig.~\ref{fig:figure 6}, we present such a generalization adopting an intercell coupling scheme. From our 1D examples we already know that in the effective Hamiltonian, it is the $v$ and $t$ hopping amplitudes that will be exponentially suppressed due to the intercell 
coupling to baths. In addition, there will be bath-induced fermion-fermion interactions between electrons occupying neighboring unit cells. 
If one introduces an open boundary condition, it can be readily shown that the $t$ hopping amplitudes along the two edges will not be affected by the baths. Therefore, an edge modes will persist along the entire boundary. Other dissipative, higher dimensional, or spin-full models exhibiting symmetry-protected topological phases can be studied via the RCPT method. 



\section*{Acknowledgments}
We acknowledge discussions with Nicholas Anto-Sztrikacs, Yuxuan Zhang, Marlon Brenes, Jakub Garwola, and Malcolm Kennett. DS acknowledges support from an NSERC Discovery Grant and the Canada Research Chair program. KA acknowledges support of the Material Sciences and
Engineering Division, Basic Energy Sciences, Office of Science, US-DOE.
\appendix
\begin{widetext}
\section{Details on the derivation of the effective Hamiltonian}
\label{sec: Details}

Here, we go through the details of deriving the effective SSH Hamiltonian via the RCPT machinery under the intracell coupling scheme. The effective Hamiltonian in the intercell coupling model can be trivially obtained from the intracell coupling scheme result. We begin with the following total Hamiltonian 
\begin{equation}
    \hat{H} = \hat{H}_\text{SSH} +\sum^L_{n=1}\hat{S}^\text{intra}_n\sum_kt_k\left(\hat{c}^\dagger_{n,k}+\hat{c}_{n,k}\right)+\sum_{n,k}\nu_k\hat{c}^\dagger_{n,k}\hat{c}_{n,k},
\end{equation}
where $\hat{H}_\text{SSH}$ is the periodic SSH model. It is given in terms of the tunneling parameters $v$ and $w$ according to
\begin{equation}
    \hat{H}_\text{SSH} =-v\sum^L_{i=1}\left(\hat{d}^\dagger_{i,A}\hat{d}_{i,B}+\hat{d}^\dagger_{i,B}\hat{d}_{i,A}\right)-w\sum^{L-1}_{i=1}\left(\hat{d}^\dagger_{i,B}\hat{d}_{i+1,A}+\hat{d}^\dagger_{i+1,A}\hat{d}_{i,B}\right)-w\left(\hat{d}^\dagger_{L,B}\hat{d}_{1,A}+\hat{d}^\dagger_{1,A}\hat{d}_{L,B}\right). 
    \label{eq:HSSHA}
\end{equation}
The system operator that couples to the $n$th bath is given by $\hat{S}^\text{intra}_n = \hat{d}^\dagger_{n,A}\hat{d}_{n,B}+\hat{d}^\dagger_{n,B}\hat{d}_{n,A}$. 
For simplicity, we drop the superscript ``intra" from here on. Recall that $n$ is the bath index, which is identical to the unit cell index $i$ for the intracell coupling scheme. We therefore use them interchangeably where appropriate. 
For clarity, we remind the reader of the RC Hamiltonian obtained prior to the polaron transform,
\begin{equation}
    \hat{H}_\text{RC} = \hat{H}_\text{SSH}+\lambda\sum^L_{n=1}\hat{S}^\text{intra}_n\left(\hat{a}^\dagger_n+\hat{a}_n\right)+\sum^L_{n=1}\Omega\hat{a}^\dagger_n\hat{a}_n+\sum_{n}\left(\hat{a}^\dagger_n+\hat{a}_n\right)\sum_{k}f_k\left(\hat{b}^\dagger_{n,k}+\hat{b}_{n,k}\right)+\sum_{n,k}\omega_k\hat{b}^\dagger_{n,k}\hat{b}_{n,k}.
\end{equation}
Here, $\{\hat{a}_n\}$ are the RC operators extracted from each local bath, with bath index $n$. The effective system Hamiltonian is given by consecutive polaron transformations applied onto the above Hamiltonian, followed by a truncation of each of the RC modes to their ground state. That is, 
\begin{equation}
\label{eq: effh intra}
    \hat{H}^\text{intra}_\text{SSH,eff} = \bra{0_1,\dots,0_L}\prod^L_{n=1}\left(\hat{U}_{n}\right)\hat{H}_\text{SSH}\prod^L_{n=1}\left(\hat{U}^\dagger_n\right)\ket{0_1,\dots,0_L}-\frac{\lambda^2}{\Omega}\sum^L_{n=1}\hat{S}^2_n,
\end{equation}
see Eq. (\ref{eq:HSeff}).
Here, $\hat{U}_n=\exp\left[\frac{\lambda}{\Omega}\hat{S}_n(\hat{a}^\dagger_n-\hat{a}_n)\right]$ is the polaron transform unitary acting on the system and the RC mode. In order for the RCPT treatment to work without ambiguity, the individual polaron transforms need to commute with one another. This is indeed the case for our choice of $\hat{S}_n$. The bosonic operators trivially commute with one another for different site index $n$ and one can easily show that $[\hat{S}_n,\hat{S}_m]=0$ for $n\neq m$ since the hopping Hamiltonians commute with another if they do not share lattice site(s). 
Furthermore, the truncation to the lowest RC mode commutes with the polaron unitary on the different subspace. This means we can write Eq.~\eqref{eq: effh intra} in a more convenient form to analyze
\begin{equation}
\label{eq: intra-cell coupling effh def}
    \hat{H}^\text{intra}_\text{SSH,eff} = \bra{0_L}\hat{U}_L\bra{0_{L-1}}\hat{U}_{L-1}\dots\bra{0_2}\hat{U}_2\bra{0_1}\hat{U}_1\hat{H}_\text{SSH}\hat{U}^\dagger_1\ket{0_1}\hat{U}^\dagger_2\ket{0_2}\dots \hat{U}^\dagger_{L-1}\ket{0_{L-1}}\hat{U}^\dagger_L\ket{0_L}-\frac{\lambda^2}{\Omega}\sum^L_{n=1}\hat{S}^2_n.
    \end{equation}
 The polaron transform and the associated truncation can be performed in any order. Therefore, we will demonstrate the effect of a single transform followed by a  truncation on an arbitrary bath index $k$. That is
    \begin{equation}
        \bra{0_k}\hat{U}_k\hat{H}_\text{SSH}\hat{U}^\dagger_k\ket{0_k}.
    \end{equation}
    Explicitly writing down the polaron transform, we have
    \begin{equation}
        \hat{U}_k\hat{H}_\text{SSH}\hat{U}^\dagger_k = \hat{H}_\text{SSH}+\frac{\lambda}{\Omega}\hat{A}_k\left[\hat{S}_k,\hat{H}_\text{SSH}\right]+\frac{1}{2!}\left(\frac{\lambda}{\Omega}\right)^2\hat{A}^2_k\left[\hat{S}_k,\left[\hat{S}_k,\hat{H}_\text{SSH}\right]\right]+\frac{1}{3!}\left(\frac{\lambda}{\Omega}\right)^2\hat{A}^3_k\left[\hat{S}^2_k,\left[\hat{S}_k,\left[\hat{S}_k,\hat{H}_\text{SSH}\right]\right]\right]+\dots,
    \end{equation}
    where $\hat{A}_k=\hat{a}^\dagger_k-\hat{a}_k$. 
    Let us now consecutively compute the nested commutators. First, we evaluate
    \begin{equation}
        \left[\hat{S}_k,\hat{H}_\text{SSH}\right] = \left[\hat{d}^\dagger_{k,A}\hat{d}_{k,B}+\hat{d}^\dagger_{k,B}\hat{d}_{k,A}, -w\sum^{L-1}_{i=1}\left(\hat{d}^\dagger_{i,B}\hat{d}_{i+1,A}+\hat{d}^\dagger_{i+1,A}\hat{d}_{i,B}\right)-w\left(\hat{d}^\dagger_{L,B}\hat{d}_{1,A}+\hat{d}^\dagger_{1,A}\hat{d}_{L,B}\right) \right].
    \end{equation}
    The intracell hopping terms of the Hamiltonian with the hopping amplitudes $v$ immediately commutes with our choice of $\hat{S}_k$. Therefore, we only need to consider the $w$ hopping in the commutators. Out of these $w$ hopping terms, only those that share the same fermion sites will have a non-zero commutator (nearest-neighbour to the left and right). Hence, the commutator relation reduces to 
    \begin{equation}
    \begin{aligned}
        \left[\hat{S}_k,\hat{H}_\text{SSH}\right] =&\left[\hat{d}^\dagger_{k,A}\hat{d}_{k,B}+\hat{d}^\dagger_{k,B}\hat{d}_{k,A},-w\left(\hat{d}^\dagger_{k,B}\hat{d}_{k+1,A}+\hat{d}^\dagger_{k+1,A}\hat{d}_{k,B}\right)-w\left(\hat{d}^\dagger_{k-1,B}\hat{d}_{k,A}+\hat{d}^\dagger_{k,A}\hat{d}_{k-1,B}\right)\right]\\
        =&-w\left[\hat{d}^\dagger_{k,A}\hat{d}_{k,B}+\hat{d}^\dagger_{k,B}\hat{d}_{k,A},\hat{d}^\dagger_{k,B}\hat{d}_{k+1,A}+\hat{d}^\dagger_{k+1,A}\hat{d}_{k,B}\right]-w\left[\hat{d}^\dagger_{k,A}\hat{d}_{k,B}+\hat{d}^\dagger_{k,B}\hat{d}_{k,A},\hat{d}^\dagger_{k-1,B}\hat{d}_{k,A}+\hat{d}^\dagger_{k,A}\hat{d}_{k-1,B}\right].
        \end{aligned}
    \end{equation}
    Using a known commutation relation for fermionic nearest-neighbour hopping terms, 
    $[\hat{d}^\dagger_i\hat{d}_j+\hat{d}^\dagger_j\hat{d}_i,\hat{d}^\dagger_j\hat{d}_k+\hat{d}^\dagger_k\hat{d}_j]=\hat{d}^\dagger_i\hat{d}_k-\hat{d}^\dagger_k\hat{d}_i$, the above commutation relation reduces to 
    \begin{equation}
         \left[\hat{S}_k,\hat{H}_\text{SSH}\right] =-w\left(\hat{d}^\dagger_{k,A}\hat{d}_{k+1,A}-\hat{d}^\dagger_{k+1,A}\hat{d}_{k,A}\right)-w\left(\hat{d}^\dagger_{k,B}\hat{d}_{k-1,B}-\hat{d}^\dagger_{k-1,B}\hat{d}_{k,B}\right).
    \end{equation}
    We now compute the second-order contribution, which is given by 
    \begin{equation}
    \label{eq: second order nested com }
\begin{aligned}\left[\hat{S}_k,\left[\hat{S}_k,\hat{H}_\text{SSH}\right]\right]=&\left[\hat{d}^\dagger_{k,A}\hat{d}_{k,B}+\hat{d}^\dagger_{k,B}\hat{d}_{k,A},-w\left(\hat{d}^\dagger_{k,A}\hat{d}_{k+1,A}-\hat{d}^\dagger_{k+1,A}\hat{d}_{k,A}\right)-w\left(\hat{d}^\dagger_{k,B}\hat{d}_{k-1,B}-\hat{d}^\dagger_{k-1,B}\hat{d}_{k,B}\right)\right]\\
=&-w\left[\hat{d}^\dagger_{k,A}\hat{d}_{k,B}+\hat{d}^\dagger_{k,B}\hat{d}_{k,A},\hat{d}^\dagger_{k,A}\hat{d}_{k+1,A}-\hat{d}^\dagger_{k+1,A}\hat{d}_{k,A}\right]-w\left[\hat{d}^\dagger_{k,A}\hat{d}_{k,B}+\hat{d}^\dagger_{k,B}\hat{d}_{k,A},\hat{d}^\dagger_{k,B}\hat{d}_{k-1,B}-\hat{d}^\dagger_{k-1,B}\hat{d}_{k,B}\right].
    \end{aligned}
    \end{equation}
    Let us compute the first commutator; the second commutator will have the same form,
    \begin{equation}
\begin{aligned}\left[\hat{d}^\dagger_{k,A}\hat{d}_{k,B}+\hat{d}^\dagger_{k,B}\hat{d}_{k,A},\hat{d}^\dagger_{k,A}\hat{d}_{k+1,A}-\hat{d}^\dagger_{k+1,A}\hat{d}_{k,A}\right] =&\left[\hat{d}^\dagger_{k,A}\hat{d}_{k,B},\hat{d}^\dagger_{k,A}\hat{d}_{k+1,A}\right]-\left[\hat{d}^\dagger_{k,A}\hat{d}_{k,B},\hat{d}^\dagger_{k+1,A}\hat{d}_{k,A}\right]\\
+&\left[\hat{d}^\dagger_{k,B}\hat{d}_{k,A},\hat{d}^\dagger_{k,A}\hat{d}_{k+1,A}\right]-\left[\hat{d}^\dagger_{k,B}\hat{d}_{k,A},\hat{d}^\dagger_{k+1,A}\hat{d}_{k,A}\right]
        \end{aligned}
    \end{equation}
    Using the commutation identity, $[\hat{\mathcal{O}}_1\hat{\mathcal{O}}_2,\hat{\mathcal{O}}_3]=\hat{\mathcal{O}}_1[\hat{\mathcal{O}}_2,\hat{\mathcal{O}}_3]+[\hat{\mathcal{O}}_1,\hat{\mathcal{O}}_3]\hat{\mathcal{O}}_2$, we can write 
      \begin{equation}
\begin{aligned}\left[\hat{d}^\dagger_{k,A}\hat{d}_{k,B}+\hat{d}^\dagger_{k,B}\hat{d}_{k,A},\hat{d}^\dagger_{k,A}\hat{d}_{k+1,A}-\hat{d}^\dagger_{k+1,A}\hat{d}_{k,A}\right] =&\hat{d}^\dagger_{k,A}\left[\hat{d}_{k,B},\hat{d}^\dagger_{k,A}\hat{d}_{k+1,A}\right]+\left[\hat{d}^\dagger_{k,A},\hat{d}^\dagger_{k,A}\hat{d}_{k+1,A}\right]\hat{d}_{k,B}\\
-&\hat{d}^\dagger_{k,A}\left[\hat{d}_{k,B},\hat{d}^\dagger_{k+1,A}\hat{d}_{k,A}\right]-\left[\hat{d}^\dagger_{k,A},\hat{d}^\dagger_{k+1,A}\hat{d}_{k,A}\right]\hat{d}_{k,B}\\
+&\hat{d}^\dagger_{k,B}\left[\hat{d}_{k,A},\hat{d}^\dagger_{k,A}\hat{d}_{k+1,A}\right]+\left[\hat{d}^\dagger_{k,B},\hat{d}^\dagger_{k,A}\hat{d}_{k+1,A}\right]\hat{d}_{k,A}\\
-&\hat{d}^\dagger_{k,B}\left[\hat{d}_{k,A},\hat{d}^\dagger_{k+1,A}\hat{d}_{k,A}\right]-\left[\hat{d}^\dagger_{k,B},\hat{d}^\dagger_{k+1,A}\hat{d}_{k,A}\right]\hat{d}_{k,A}.
        \end{aligned}
    \end{equation}
    The only non-zero contributions will be the commutators involving creation and annihilation operators with the same index. Therefore, the only surviving terms are fourth and fifth contributions. Hence,
    \begin{equation}
\begin{aligned}\left[\hat{d}^\dagger_{k,A}\hat{d}_{k,B}+\hat{d}^\dagger_{k,B}\hat{d}_{k,A},\hat{d}^\dagger_{k,A}\hat{d}_{k+1,A}-\hat{d}^\dagger_{k+1,A}\hat{d}_{k,A}\right] =&\hat{d}^\dagger_{k,B}\left[\hat{d}_{k,A},\hat{d}^\dagger_{k,A}\hat{d}_{k+1,A}\right]-\left[\hat{d}^\dagger_{k,A},\hat{d}^\dagger_{k+1,A}\hat{d}_{k,A}\right]\hat{d}_{k,B}\\
=&\hat{d}^\dagger_{k,B}\left(\hat{d}_{k,A}\hat{d}^\dagger_{k,A}\hat{d}_{k+1,A}-\hat{d}^\dagger_{k,A}\hat{d}_{k+1,A}\hat{d}_{k,A}\right)\\
-&\left(\hat{d}^\dagger_{k,A}\hat{d}^\dagger_{k+1,A}\hat{d}_{k,A}-\hat{d}^\dagger_{k+1,A}\hat{d}_{k,A}\hat{d}^\dagger_{k,A}\right)\hat{d}_{k,B}\\
=&\hat{d}^\dagger_{k,B}\hat{d}_{k+1,A}\{\hat{d}_{k,A},\hat{d}^\dagger_{k_A}\}+\{\hat{d}^\dagger_{k,A},\hat{d}_{k,A}\}\hat{d}^\dagger_{k+1,A}\hat{d}_{k,B}\\
=&\hat{d}^\dagger_{k,B}\hat{d}_{k+1,A}+\hat{d}^\dagger_{k+1,A}\hat{d}_{k,B}
\end{aligned}
    \end{equation}
Due to symmetry, the second commutator from Eq.~\eqref{eq: second order nested com } will be
\begin{equation}
    \left[\hat{d}^\dagger_{k,A}\hat{d}_{k,B}+\hat{d}^\dagger_{k,B}\hat{d}_{k,A},\hat{d}^\dagger_{k,B}\hat{d}_{k-1,B}-\hat{d}^\dagger_{k-1,B}\hat{d}_{k,B}\right]=\hat{d}^\dagger_{k-1,B}\hat{d}_{k,A}+\hat{d}^\dagger_{k,A}\hat{d}_{k-1,B}
\end{equation}
We therefore have
\begin{equation} \left[\hat{S}_k\left[\hat{S}_k,\hat{H}_\text{SSH}\right]\right] = -w\left(\hat{d}^\dagger_{k,B}\hat{d}_{k+1,A}+\hat{d}^\dagger_{k+1,A}\hat{d}_{k,B}\right)-w\left(\hat{d}^\dagger_{k-1,B}\hat{d}_{k,A}+\hat{d}^\dagger_{k,A}\hat{d}_{k-1,B}\right).
\end{equation}
That is, we recover the same hopping Hamiltonians that is nearest neighbors to the left and to the right of our selected unit cell $k$. 
Iterating,  we are able to write 
\begin{equation}
\label{eq: polaron transform intra cell before truncation}
\begin{aligned}
    \hat{U}_k\hat{H}_\text{SSH}\hat{U}^\dagger_k = &\hat{H}_\text{SSH}+w\left(\hat{d}^\dagger_{k,B}\hat{d}_{k+1,A}+\hat{d}^\dagger_{k+1,A}\hat{d}_{k,B}+\hat{d}^\dagger_{k-1,B}\hat{d}_{k,A}+\hat{d}^\dagger_{k,A}\hat{d}_{k-1,B}\right)\\
    -&w\cosh\Big(\frac{\lambda}{\Omega}\hat{A}_k\Big)\left(\hat{d}^\dagger_{k,B}\hat{d}_{k+1,A}+\hat{d}^\dagger_{k+1,A}\hat{d}_{k,B}+\hat{d}^\dagger_{k-1,B}\hat{d}_{k,A}+\hat{d}^\dagger_{k,A}\hat{d}_{k-1,B}\right)\\
    -&w\sinh\Big(\frac{\lambda}{\Omega}\hat{A}_k\Big)\left(\hat{d}^\dagger_{k,A}\hat{d}_{k+1,A}-\hat{d}^\dagger_{k+1,A}\hat{d}_{k,A}
+\hat{d}^\dagger_{k,B}\hat{d}_{k-1,B}-\hat{d}^\dagger_{k-1,B}\hat{d}_{k,B}\right).
\end{aligned}
\end{equation}
That is, the polaron transform replaces the hopping Hamiltonian (second term of Eq.~\eqref{eq: polaron transform intra cell before truncation}) with a bath-operator dressed one (third term of Eq.~\eqref{eq: polaron transform intra cell before truncation}) and further introduces a next-nearest neighbour hopping via the last term of Eq.~\eqref{eq: polaron transform intra cell before truncation}. 

Next, we project the Hamiltonian into the ground state of the RC mode. This allows us to recover the system Hamiltonian, with renormalized parameters. That is,
\begin{equation}
    \begin{aligned}\bra{0_k}\hat{U}_k\hat{H}_\text{SSH}\hat{U}^\dagger_k\ket{0_k} =& \hat{H}_\text{SSH}+w\left(\hat{d}^\dagger_{k,B}\hat{d}_{k+1,A}+\hat{d}^\dagger_{k+1,A}\hat{d}_{k,B}+\hat{d}^\dagger_{k-1,B}\hat{d}_{k,A}+\hat{d}^\dagger_{k,A}\hat{d}_{k-1,B}\right)\\
    -&w\bra{0_k}\cosh\Big(\frac{\lambda}{\Omega}\hat{A}_k\Big)\ket{0_k}\left(\hat{d}^\dagger_{k,B}\hat{d}_{k+1,A}+\hat{d}^\dagger_{k+1,A}\hat{d}_{k,B}+\hat{d}^\dagger_{k-1,B}\hat{d}_{k,A}+\hat{d}^\dagger_{k,A}\hat{d}_{k-1,B}\right)\\
    -&w\bra{0_k}\sinh\Big(\frac{\lambda}{\Omega}\hat{A}_k\Big)\ket{0_k}\left(\hat{d}^\dagger_{k,A}\hat{d}_{k+1,A}-\hat{d}^\dagger_{k+1,A}\hat{d}_{k,A}
+\hat{d}^\dagger_{k,B}\hat{d}_{k-1,B}-\hat{d}^\dagger_{k-1,B}\hat{d}_{k,B}\right).
    \end{aligned}
\end{equation}
Note that $\bra{0_k}\sinh\Big(\frac{\lambda}{\Omega}\hat{A}_k\Big)\ket{0_k}=0$ since $\hat{A}_k=\hat{a}^\dagger_k-\hat{a}_k$ and the hyperbolic sine will involve only odd powers of $\hat{A}_k$. 
We are able to evaluate $\bra{0_k}\cosh\Big(\frac{\lambda}{\Omega}\hat{A}_k\Big)\ket{0_k}$ as 
\begin{equation}
    \bra{0_k}\cosh\Big(\frac{\lambda}{\Omega}\hat{A}_k\Big)\ket{0_k} = \frac{1}{2}\bra{0_k}e^{\frac{\lambda}{\Omega}(\hat{a}^\dagger_k-\hat{a}_k)}+e^{-\frac{\lambda}{\Omega}(\hat{a}^\dagger_k-\hat{a}_k)}\ket{0_k}.
\end{equation}
Defining a displacement operator, $D(\alpha)=e^{\alpha(\hat{a}^\dagger_k-\hat{a}_k)}$, where $\alpha=\frac{\lambda}{\Omega}\hat{I}$, we get
\begin{equation}
\begin{aligned}
    \bra{0_k}\cosh\Big(\frac{\lambda}{\Omega}\hat{A}_k\Big)\ket{0_k} =& \frac{1}{2}\bra{0_k}D(\alpha)+D(-\alpha)\ket{0_k}\\
    =&\frac{1}{2}\bra{0_k}\left(e^{-\frac{|\lambda/\Omega|^2}{2}}\sum^\infty_{n=0}\frac{(\frac{\lambda}{\Omega})^2}{\sqrt{n!}}\ket{n_k}+e^{-\frac{|\lambda/\Omega|^2}{2}}\sum^\infty_{n=0}\frac{(-\frac{\lambda}{\Omega})^2}{\sqrt{n!}}\ket{n_k}\right)\\
    =&\frac{1}{2}\bra{0_k}e^{-\frac{\lambda^2}{2\Omega^2}}+e^{-\frac{\lambda^2}{2\Omega^2}}\ket{0_k}\\
    =&e^{-\frac{\lambda^2}{2\Omega^2}}.
    \end{aligned}
\end{equation}
Therefore, we finally have
\begin{equation}
\label{eq: single polaron transform intra}
\begin{aligned}\bra{0_k}\hat{U}_k\hat{H}_\text{SSH}\hat{U}^\dagger_k\ket{0_k} =& \hat{H}_\text{SSH}+w\left(\hat{d}^\dagger_{k,B}\hat{d}_{k+1,A}+\hat{d}^\dagger_{k+1,A}\hat{d}_{k,B}+\hat{d}^\dagger_{k-1,B}\hat{d}_{k,A}+\hat{d}^\dagger_{k,A}\hat{d}_{k-1,B}\right)\\
    -&we^{-\frac{\lambda^2}{2\Omega^2}}\left(\hat{d}^\dagger_{k,B}\hat{d}_{k+1,A}+\hat{d}^\dagger_{k+1,A}\hat{d}_{k,B}+\hat{d}^\dagger_{k-1,B}\hat{d}_{k,A}+\hat{d}^\dagger_{k,A}\hat{d}_{k-1,B}\right)\\
    \end{aligned}
\end{equation}
Once we perform all the consecutive polaron transforms along with the  truncation of all RCs for all the intracell hopping terms, we end up with
\begin{equation}
    \bra{0_L}\hat{U}_L\bra{0_{L-1}}\hat{U}_{L-1}\dots\bra{0_2}\hat{U}_2\bra{0_1}\hat{U}_1\hat{H}_\text{SSH}\hat{U}^\dagger_1\ket{0_1}\hat{U}^\dagger_2\ket{0_2}\dots \hat{U}^\dagger_{L-1}\ket{0_{L-1}}\hat{U}^\dagger_L\ket{0_L} = \hat{H}_\text{SSH}(v,\tilde{w}),
\end{equation}
where $\tilde{w}=we^{-\frac{\lambda^2}{\Omega^2}}$. 
The additional factor of two in the exponent, compared to the suppression written in Eq.~\eqref{eq: single polaron transform intra}, arises due to each intercell hopping Hamiltonian sharing two neighboring intracell hopping terms. 

We now examine the bath-induced interaction term given by the last term in Eq.~\eqref{eq: intra-cell coupling effh def}
\begin{equation}
\begin{aligned}
    -\frac{\lambda^2}{\Omega}\sum^L_{n=1}\hat{S}^2_n =& -\frac{\lambda^2}{\Omega}\sum^L_{n=1}\left(\hat{d}^\dagger_{n,A}\hat{d}_{n,B}+\hat{d}^\dagger_{n,B}\hat{d}_{n,A}\right)^2\\
    =&-\frac{\lambda^2}{\Omega}\sum^L_{n=1}\left(\hat{d}^\dagger_{n,A}\hat{d}_{n,B}\hat{d}^\dagger_{n,B}\hat{d}_{n,A}+\hat{d}^\dagger_{n,B}\hat{d}_{n,A}\hat{d}^\dagger_{n,A}\hat{d}_{n,B}\right)\\
    =&-\frac{\lambda^2}{\Omega}\sum^L_{n=1}\left(\hat{d}^\dagger_{n,A}\hat{d}_{n,A}\left(1-\hat{d}^\dagger_{n,B}\hat{d}_{n,B}\right)+\hat{d}^\dagger_{n,B}\hat{d}_{n,B}\left(1-\hat{d}^\dagger_{n,A}\hat{d}_{n,A}\right)\right)\\
    =&-\frac{\lambda^2}{\Omega}\sum^L_{n=1}\left(\hat{d}^\dagger_{n,A}\hat{d}_{n,A}-\hat{d}^\dagger_{n,B}\hat{d}_{n,B}\right)^2.
    \end{aligned}
\end{equation}
Putting together the tunneling terms with the many-body interaction, the system Hamiltonian is given by 
\begin{equation}
    \hat{H}^\text{intra}_\text{SSH,eff} = \hat{H}_\text{SSH}(v,\tilde{w})-\frac{\lambda^2}{\Omega}\sum^L_{i=1}\left(\hat{n}_{i,A}-\hat{n}_{i,B}\right)^2
\end{equation}
By symmetry, in the case of the intercell coupling scheme, the $v$ hopping is exponentially suppressed, instead of $w$, and the baht-induced interaction term develops between $\hat{n}_{i+1,A}$ and $\hat{n}_{i,B}$. That is,
\begin{equation}
    \hat{H}^\text{inter}_\text{SSH,eff} = \hat{H}_\text{SSH}(\tilde{v},w)-\frac{\lambda^2}{\Omega}\sum^L_{i=1}\left(\hat{n}_{i+1,A}-\hat{n}_{i,B}\right)^2-\frac{\lambda^2}{\Omega}(\hat{n}_{1,A}-\hat{n}_{L,B})^2,
\end{equation}
where $\tilde{v}=ve^{-\frac{\lambda^2}{\Omega^2}}$.

\section{Rice-Mele model: Staggered potential}
\label{sec: staggered potential}

Let us examine the effect of the intracell coupling scheme to the staggered potential terms. Consider the following Hamiltonian
\begin{equation}
    \hat{H} =   \sum^L_{i=1}\hat{H}_{\text{SP},i}(u)   +H_{\text{SSH}}(v,w)
    +\sum^L_{n=1}\hat{S}^\text{intra}_n\sum_kt_k\left(\hat{c}^\dagger_{n,k}+\hat{c}_{n,k}\right)+\sum_{n,k}\nu_k\hat{c}^\dagger_{n,k}\hat{c},_{n,k},
\end{equation}
where $ \hat{H}_{\text{SP},i}(u)= u(\hat{d}^\dagger_{i,A}\hat{d}_{i,A}-\hat{d}^\dagger_{i,B}\hat{d}_{i,B})$, the SSH chain Hamiltonian is as given by Eq. (\ref{eq:HSSHA}),  $\hat{S}^\text{intra}_n=\hat{d}^\dagger_{n,A}\hat{d}_{n,B}+h.c.$. 
Again, we will interchangeably use the unit-cell index $i$ and the bath index $n$. Since we already know that this coupling scheme will generate many-body interaction terms given by $\hat{H}^\text{intra}_\text{MB}=-\frac{\lambda^2}{\Omega}\sum^L_{i=1}(\hat{d}^\dagger_{i,A}\hat{d}_{i,A}-\hat{d}^\dagger_{i,B}\hat{d}_{i,B})^2$, we just need to work out how the $u$ parameter is renormalized under the intracell coupling scheme. That is, we need to compute
\begin{equation}
\bra{0_1,\dots,0_L}\prod^L_{n=1}\left(\hat{U}_{n}\right)\sum^L_{i=1}\hat{H}_{\text{SP},i}(u)\prod^L_{n=1}\left(\hat{U}^\dagger_n\right)\ket{0_1,\dots,0_L},
\end{equation}
where $\hat{U}_n=\exp\left[\frac{\lambda}{\Omega}\hat{S}_n(\hat{a}^\dagger_n-\hat{a}_n)\right]$. Again, since the hopping Hamiltonians commute with one another, we are able to more conveniently write the above expression as 
\begin{equation}
    \bra{0_L}\hat{U}_L\bra{0_{L-1}}\hat{U}_{L-1}\dots\bra{0_2}\hat{U}_2\bra{0_1}\hat{U}_1\sum^L_{i=1}\hat{H}_{\text{SP},i}(u)\hat{U}^\dagger_1\ket{0_1}\hat{U}^\dagger_2\ket{0_2}\dots \hat{U}^\dagger_{L-1}\ket{0_{L-1}}\hat{U}^\dagger_L\ket{0_L}.
\end{equation}
Since the unitary transforms only affect fermions that are sharing the same lattice sites as the hopping Hamiltonian in the exponential, we only need to focus on a sample unit-cell $n$. That is,
\begin{equation}
\begin{aligned}\bra{0_n}\hat{U}^\dagger_n\hat{H}_{\text{SP},n}\hat{U}_n\ket{0_n} 
= \hat{H}_{\text{SP},n}
+\frac{\lambda}{\Omega}\hat{A}_n\left[\hat{S}_n,\hat{H}_{\text{SP},n}\right]
+\frac{1}{2!}\left(\frac{\lambda}{\Omega}\right)^2\hat{A}^2_n\left[\hat{S}_n,\left[\hat{S}_n,\hat{H}_{\text{SP},n}\right]\right]+\dots,
\end{aligned}
\end{equation}
where $\hat{A}_n=\hat{a}^\dagger_n-\hat{a}_n$. Systematically computing the nested commutators, we first find after some algebra
%
\begin{equation}
    \left[\hat{S}_n,\hat{H}_{\text{SP},n}\right] = 2u\left(\hat{d}^\dagger_{n,B}\hat{d}_{n,A}-\hat{d}^\dagger_{n,A}\hat{d}_{n,B}\right).
\end{equation}
Next, for the first nested commutator, we get
\begin{equation}
    \left[\hat{S}_n,\left[\hat{S}_n,\hat{H}_{\text{SP},n}\right]\right] = 4u\left(\hat{d}^\dagger_{n,A}\hat{d}_{n,A}-\hat{d}_{n,B}\hat{d}_{n,B}\right),
\end{equation}
for which the pattern continues. We therefore have
\begin{equation}
\bra{0_n}\hat{U}^\dagger_n\hat{H}_{\text{SP},n}\hat{U}_n\ket{0_n} =\bra{0_n}\hat{H}_{\text{SP},n}
+\frac{2\lambda}{\Omega}u\hat{A}_n\left(\hat{d}^\dagger_{n,B}\hat{d}_{n,A}-\hat{d}^\dagger_{n,A}\hat{d}_{n,B} \right)+\frac{1}{2!}\left(\frac{2\lambda}{\Omega}\right)^2\hat{H}_{\text{SP},n}+\dots\ket{0_n} .
\end{equation}
Since all the odd 
terms are zero after projecting to the ground state of the reaction coordinates we are left with
\begin{equation}
\bra{0_n}\hat{U}^\dagger_n\hat{H}_{\text{SP},n}\hat{U}_n\ket{0_n} = u\bra{0_n}\cosh\Big[\frac{2\lambda}{\Omega}\left(\hat{a}^\dagger_n-\hat{a}_n\right)\Big]\ket{0_n}\left(\hat{d}^\dagger_{n,A}\hat{d}_{n,A}-\hat{d}_{n,B}\hat{d}_{n,B}\right).
\end{equation}
Utilizing the displacement operator, $D(\alpha)=e^{\alpha(\hat{a}^\dagger_n-\hat{a}_n)}$, where $\alpha=\frac{2\lambda}{\Omega}\hat{I}$, we find
\begin{equation}
\begin{aligned}
    \bra{0_n}\cosh\Big(\frac{2\lambda}{\Omega}\hat{A}_n\Big)\ket{0_n} =& \frac{1}{2}\bra{0_n}D(\alpha)+D(-\alpha)\ket{0_n}\\
    =&\frac{1}{2}\bra{0_n}\left(e^{-\frac{|2\lambda/\Omega|^2}{2}}\sum^\infty_{n=0}\frac{(\frac{2\lambda}{\Omega})^n}{\sqrt{n!}}\ket{n_n}+e^{-\frac{|2\lambda/\Omega|^2}{2}}\sum^\infty_{n=0}\frac{(-\frac{2\lambda}{\Omega})^n}{\sqrt{n!}}\ket{n_n}\right)\\
    =&\frac{1}{2}\bra{0_n}e^{-\frac{2\lambda^2}{\Omega^2}}+e^{-\frac{2\lambda^2}{\Omega^2}}\ket{0_n}\\
    =&e^{-\frac{2\lambda^2}{\Omega^2}}.
    \end{aligned}
\end{equation}
Therefore, the staggered potential is suppressed by the factor of $e^{-\frac{2\lambda^2}{\Omega^2}}$. 
The effective system Hamiltonian of the Rice-Mele model would then be
\begin{equation}
    \hat{H}^\text{intra,eff}_\text{SP}(u) = \hat{H}_\text{SP}(\tilde{u})+H_{\text{SSH}}(v, \tilde w)+\hat{H}^\text{intra}_\text{MB},
\end{equation}
where $\tilde{u}=ue^{-\frac{2\lambda^2}{\Omega^2}}$. 

If we were to have more general starting chemical potential say $\mu_A$ for the $A$ sites and $\mu_B$ for the $B$ sites, the original chemical potentials will be renormalized to 
\begin{equation}
\begin{aligned}
    \tilde{\mu}_A =&\left[\mu_A\cosh\Big(\frac{\lambda^2}{\Omega^2}\Big)+\mu_B\sinh\Big(\frac{\lambda^2}{\Omega^2} \Big)\right]e^{-\frac{\lambda^2}{\Omega^2}}\\
    \tilde{\mu}_B =& \left[\mu_B\cosh\Big(\frac{\lambda^2}{\Omega^2}\Big)+\mu_A\sinh\Big(\frac{\lambda^2}{\Omega^2} \Big)\right]e^{-\frac{\lambda^2}{\Omega^2}}
    \end{aligned}.
\end{equation}
For $\mu_A=u$ and $\mu_B=-u$, the above formula indeed goes to the total suppression of $e^{-\frac{2\lambda^2}{\Omega^2}}$.


\section{Inadequacy of the MF treatment of the bath-induced interactions}
\label{sec: MF theory}

In this appendix, we demonstrate that a mean-field (MF) approximation to the bath-induced many-body interaction terms will lead to an \textit{incorrect} phase diagram. One might initially argue that the dissipative schemes presented in the main text does not strictly depict one-dimensional scenarios. That is, as we introduce bosonic baths on each unit-cell (or unit-cell boundary) of the SSH chain, it is tempting to imagine that there might exist a long-range order, even at finite temperature. For instance, given the form of the bath-induced many-body interaction for the intracell coupling scheme, $\hat{H}^\text{intra}_\text{MB}=-\frac{\lambda^2}{\Omega}\sum^L_{i=1}(\hat{n}_{i,A}-\hat{n}_{i,B})^2$, a charge density wave seems to be a possible instability. This is because it is energetically advantageous to occupy solely $A$ sites or $B$ sites. 

Let us now go through the standard MF procedure for the intracell coupling scheme. The intercell coupling scheme result can be trivially obtained afterwards. First, we note that the interacting Hamiltonian described by the last term of Eq.~\eqref{eq: intra EFFH} can be approximated in the following way: 
If we define the operator, $\hat{\chi}_{i,AB}=\hat{n}_{i,A}-\hat{n}_{i,B}$, it allows us to write the interaction term as
 $-\frac{\lambda^2}{\Omega}\sum^L_{i=1}\hat{\chi}^2_{i,AB}$. Then, we decompose $\hat{\chi}_{i,AB}=\langle\hat{\chi}_{i,AB}\rangle_\text{MF}+\delta \hat{\chi}_{i,AB}$ where the $\chi_0\equiv\langle\hat{\chi}_{i,AB}\rangle_\text{MF}$ and $\delta \hat{\chi}_{i,AB}$ correspond to a CDW order parameter, and  a small deviation from it respectively. This leads to 
\begin{equation}
\begin{aligned}
    \chi^2_{i,AB} 
    =&\chi_0^2+2\chi_0\delta \hat{\chi}_{i,AB}+\left(\delta\hat{\chi}_{i,AB}\right)^2\\
    \simeq& \chi_0^2+2\chi_0\left(\hat{\chi}_{i,AB}-\chi_0\right)\\
    =&2\chi_0\hat{\chi}_{i,AB}-\chi_0^2.
    \end{aligned}
\end{equation}
The MF Hamiltonian we analyze is therefore given as 
\begin{equation}
\label{eq: intra EFFH MF}
\begin{aligned}
    \hat{H}^\text{intra,MF}_\text{SSH,eff} =\hat{H}_\text{SSH}(v,\tilde{w})
    -\frac{\lambda^2}{\Omega}\sum^L_{i=1}\left(2\chi_0\left(\hat{n}_{i,A}-\hat{n}_{i,B}\right)-\chi_0^2\right).
    \end{aligned}
\end{equation}
Similarly, performing the MF approximation on the intercell coupling Hamiltonian described by Eq.~\eqref{eq: inter EFFH} leads to 
\begin{equation}
\label{eq: inter EFFH MF}
    \hat{H}^\text{inter,MF}_\text{SSH,eff} = \hat{H}_\text{SSH}(\tilde{v},w)-\frac{\lambda^2}{\Omega}\sum^{L-1}_{i=1}\left(2\chi_0\left(\hat{n}_{i+1,A}-\hat{n}_{i,B}\right)-\chi_0^2\right).
\end{equation}
That is, a staggered potential emerges, suggesting a possible CDW order by breaking the sub-lattice symmetry. Already at this point, one should be alarmed at the final form of the effective Hamiltonian. The sub-lattice symmetry is crucial in defining the two topological phases. However, from the MF Hamiltonians above, if the coupling strength is above some critical value $\lambda_c$, the chain will develop a CDW order. 

We now apply the MF self-consistency loop to identify the order parameter $\chi_0$, which we will assume take a positive value. That is, we are breaking the sub-lattice symmetry by hand, so that the $A$ sites sit at a lower chemical potential. In addition, since we consider the SSH chain at half-filling, to correctly implement the fermion anti-commutation relation for many non-interacting fermions, we employ the following formula to compute the thermal average, 
\begin{equation}
\begin{aligned}
  \langle e^{\hat{M}}\rangle = \det(\hat{I}-\hat{f}(E_i,\mu)+e^{\hat{M}}\hat{f}(E_i,\mu)).
    \end{aligned}
\end{equation}
Here, $\hat{f}(E_i,\mu)$ is a diagonal matrix of Fermi-Dirac distributions with energy $E_i$ and chemical potential $\mu$. Namely: $\hat{f}(E_i,\mu)=\text{diag}(f(E_1,\mu),f(E_2,\mu),\dots,f(E_N,\mu))$. We note that $\hat{M}$ is a matrix representation of an observerable of interest in the energy basis. Since we have to compute $\chi_0=\langle \hat{\chi}_{i,AB} \rangle_\text{MF}$, which is not raised to the power of an exponential, we instead compute
\begin{equation}
\label{eq: determinant formula}  
    \chi = \lim_{\xi\rightarrow 0}\frac{d}{d\xi } \det(\hat{I}-\hat{f}(E_i,\mu)+e^{\xi\hat{\chi}^E_{i,AB}}\hat{f}(E_i,\mu))
\end{equation}
where $\hat{\chi}^E_{i,AB}=\hat{U}^\dagger\hat{\chi}_{i,AB}\hat{U}$. 
We have defined $\hat{U}$ as a unitary transformation diagonalizing $\hat{H}^\text{intra,MF}_\text{SSH,eff}=\hat{U}\hat{E}\hat{U}^\dagger$ with $\hat{E}=\text{diag}(E_1,E_2,\dots,E_N)$. 
With periodic boundary conditions, all the unit cells are identical to one another. We have therefore restricted the order parameter to the first unit cell ($i=1$). The differentiation w.r.t. $\xi$ is performed numerically, which shows excellent convergence. The determinant formula (Eq.~\eqref{eq: determinant formula}) can also handle the zero temperature limit by first finding the appropriate $\mu$ at half-filling, which is simply taken to be the average value of the middle point in the energy spectra. Then, each diagonal element of  $\hat{f}(E_i,\mu)$ is either $1$ or $0$ if $E_i<\mu$ or $E_i>\mu$, respectively. At finite temperature, we identify the correct $\mu$ at half-filling by setting the following constraint, $L=\sum^{N}_{i=1}f(E_i,\mu)$.

Once we identify the correct MF Hamiltonian and its parameter, we compute the EGP.  To correctly incorporate the  fermion anti-commutation relation, we use the following formula to extract the EGP~\cite{Bardyn_2018},
\begin{equation}
    \phi_E = \Im \ln \det[\hat{I}-f(\hat{G})+f(\hat{G})\hat{T}],
\end{equation}
where $f(\hat{G})=(e^{\hat{G}}+\hat{I})^{-1}$ and $\hat{G}=\beta(\hat{H}^\text{intra/inter}_\text{SSH,eff}-\mu\hat{N})$ with $\hat{N}=\sum^L_{i=1}(\hat{n}_{i,A}+\hat{n}_{i,B})$. 

In Fig.~\ref{fig:egp}, we plot the EGP for $L=25$, $w=1$, $\Omega=10$, at $T=0.1$ and $0.5$. We find that if $\lambda$ exceeds some critical $\lambda_c$, the chain develops a CDW order, where the EGP takes a value close to zero. The topological phase boundary that separates TI and BI here is simply given by the relative ratio of the effective $v$ and $w$ parameters. That is, it should be identical to that of the ED results presented in the main text, but where the bath-induced many-body interaction is turned off, compare Fig. \ref{fig:egp} to \ref{fig:figure 2}.

\begin{figure*}[htpb]
\fontsize{6}{10}\selectfont
\centering
\includegraphics[width=1\columnwidth]{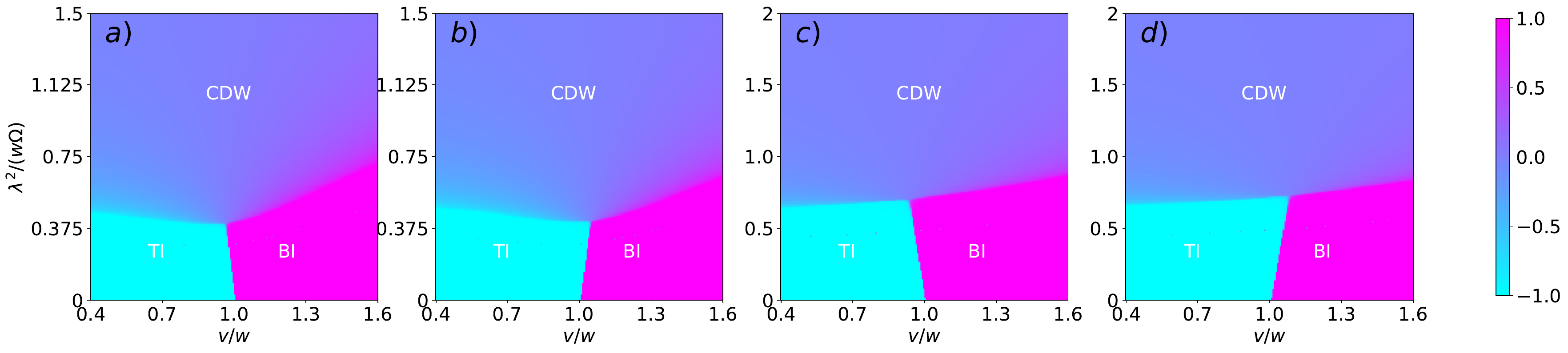}
\caption{Phase diagram of the dissipative SSH model obtain by a mean-field treatment. 
We present the EGP for intracell (a), (c) and intercell (b), (d) coupling schemes in units of $\pi/2$. 
Parameters used are $L=25$, $w=1$, $\Omega=10$.
We show results for  $T=0.1$ (a)-(b) and $T=0.5$ (c)-(d). 
Treating the bath-induced interaction terms in a MF fashion incorrectly predicts a third phase (CDW) due to the artificial breaking of the sub-lattice symmetry.  
}
\label{fig:egp}
\end{figure*}
\end{widetext}

\bibliographystyle{apsrev4-1}
\bibliography{bibliography}
\end{document}